# Efficient Plane-to-Point Energy Transfer from Supramolecular Excitonic Systems to Molecular Acceptors

Dimuthu H. Thanippuli Arachchi, Oliver M. Nix, Jianshu Cao, Moungi G. Bawendi*

*Department of Chemistry, Massachusetts Institute of Technology, 77 Massachusetts Avenue, Cambridge, Massachusetts, 02139, United States*

**Corresponding Author: mgb@mit.edu*

## ABSTRACT

Collective excitonic interactions in supramolecular assemblies offer opportunities for efficient light harvesting and directed energy transport. Here, we demonstrate efficient energy transfer from two-dimensional silica-encapsulated TDBC J-aggregate sheets to covalently attached ATTO 590 molecular acceptors. Silica encapsulation preserves the J-aggregate excitonic properties while enabling acceptor functionalization at a defined separation from the donor plane. Steady-state donor quenching and acceptor sensitization yield energy-transfer efficiencies of 75-81% in solution, while time-resolved photoluminescence measurements reveal donor lifetime shortening. We develop a generalized Förster resonance energy-transfer model for an extended planar donor coupled to localized molecular acceptors, predicting an inverse fourth-power dependence of the transfer rate on donor-acceptor separation. Using experimental spectroscopic data and estimated structural parameters, the model predicts an energy-transfer rate of $4.77 \times 10^{10}$ $s^{-1}$, approximately ten times faster than the intrinsic donor decay rate, corresponding to a theoretical energy-transfer efficiency of approximately 91%. These findings provide a framework for developing hybrid excitonic architectures for light harvesting.

In nature, photosynthetic systems comprising densely packed pigment networks exploit electronic coupling between neighboring molecules to funnel excitation energy toward localized reaction centers with quantum efficiencies exceeding 90%.[1–3] Inspired by these architectures, supramolecular excitonic materials have emerged as promising candidates for artificial light harvesting and directed energy transport, with potential applications in optoelectronic technologies.[4–6] In such systems, exciton delocalization gives rise to optical properties that differ fundamentally from those of isolated molecules, including enhanced transition dipole moments, cooperative emission,[7–12] and long range exciton transport.[6,13] Among synthetic excitonic materials, cyanine dye J-aggregates are particularly notable for their highly ordered self-assembly and strong intermolecular electronic coupling.[14,15] These aggregates exhibit narrow, red-shifted absorption bands relative to their monomers, little to no Stokes shift, superradiant emission, and large oscillator strengths due to exciton delocalization across multiple chromophores.[9,15–17] Different molecular structures and supramolecular packing geometries form diverse J-aggregate morphologies, including one-dimensional nanotubes, fibrils, ribbons, as well as two-dimensional sheet-like assemblies, providing control over their excitonic properties.[18–21] In particular, two-dimensional J-aggregate sheets formed from 5,5′,6,6′-tetrachloro-1,1′-diethyl-3,3′-di(4-sulfobutyl)-benzimidazolocarbocyanine (TDBC) exhibit strong excitonic interactions at room temperature and robust collective emission in aqueous environments.[22] Recent advances in supramolecular engineering, including silica encapsulation, matrix-assisted stabilization, and hierarchical co-assembly, have substantially improved the structural robustness, quantum yields, and chemical addressability of J-aggregates while preserving their collective excitonic properties and expanding their potential as building blocks for hybrid excitonic architectures.[13,23–28]

Despite these advances, efficiently directing excitation energy from extended excitonic assemblies to localized molecular acceptors remains challenging. Conventional Förster resonance energy transfer (FRET) describes donor and acceptor species as localized transition dipoles and therefore does not directly capture energy transfer between a spatially extended excitonic donor and a localized molecular acceptor.[29,30] In a J-aggregate donor-molecular acceptor system, excitation is delocalized across multiple electronically coupled chromophores, whereas the acceptor acts as a localized energy sink with a discrete spatial position and orientation. Consequently, donor dimensionality, exciton delocalization, donor-acceptor geometry, and separation can collectively influence energy-transfer dynamics in ways that differ from conventional point-to-point Förster transferr.[31–33] At the same time, introducing external acceptors without perturbing the supramolecular packing responsible for excitonic coupling presents an important materials design challenge. Efficient energy transfer therefore requires simultaneous preservation of the collective donor state and controlled coupling to localized acceptors.

J-aggregates have previously been integrated with molecular and inorganic acceptors through electrostatic assembly,[34–37] layer-by-layer deposition,[38] DNA templating,[39,40] and direct adsorption strategies.[41] Hybrid J-aggregate-semiconductor nanocrystal architectures have demonstrated highly efficient energy transfer and substantial enhancement of quantum-dot emission through J-aggregate-mediated light harvesting.[38,41] Related C8S3 J-aggregate-based hybrids have shown that efficient transfer can persist over nanoscale donor-acceptor separations and can be assisted by exciton transport within the aggregate.[37,42] DNA-templated J-aggregates have further enabled spatially controlled energy-transfer pathways between delocalized excitonic states and molecular fluorophores,[39,40] while molecular emitters such as DCM (4-dicyanmethylene-2-methyl-6-(p-dimethylaminostyryl)-4H-pyran) have also been sensitized by J-aggregate excitonic antennas.[43] Collectively, these studies establish the ability of J-aggregates to harvest and redistribute excitation energy in hybrid systems. However, most rely on direct association or templated assembly of the acceptor with the excitonic material, limiting independent control over donor structure, acceptor placement, and donor-acceptor separation. Energy transfer from a structurally defined two-dimensional J-aggregate donor to covalently positioned molecular acceptors therefore remains comparatively unexplored.

Here, we demonstrate energy transfer from silica-encapsulated two-dimensional TDBC J-aggregate sheets to covalently integrated ATTO 590 molecular acceptors in solution at room temperature. The silica interface preserves the collective excitonic properties of the J-aggregate sheets while providing a chemically addressable surface for positioning molecular acceptors without directly perturbing the underlying supramolecular assembly.[23] Steady state and time resolved measurements reveal efficient transfer from the delocalized excitonic donor manifold to localized molecular acceptors. We describe this extended-donor to point-acceptor interaction using a plane-to-point generalized Förster model that predicts an inverse fourth-power dependence of the energy transfer rate on donor-acceptor separation, in contrast to the inverse sixth-power dependance of conventional point-to-point energy transfer. These results establish silica-encapsulated two-dimensional J-aggregates as excitonic antennas capable of redistributing optical excitation into localized molecular states, highlighting their potential as versatile building blocks for hybrid excitonic architectures, and provide a framework for understanding resonance energy transfer in multichromophoric systems.

## RESULTS AND DISCUSSION

### *Covalent integration of ATTO 590 acceptors onto silica-encapsulated TDBC J-aggregate donors*

To construct the donor-acceptor assemblies, TDBC J-aggregate sheets were first encapsulated in a silica shell to improve structural stability, provide a surface for further functionalization, and maintain a defined separation between the J-aggregate donor and the molecular acceptor. This architecture therefore allows acceptor molecules to be introduced without direct chemical modification of the underlying TDBC assembly, while retaining control over the interfacial chemistry and donor-acceptor separation. Silica-encapsulated TDBC J-aggregates were prepared using a modified two-step in situ sol-gel procedure adapted from previously reported methods.[23–25] The overall synthetic strategy is illustrated in Fig.1 and consists of three steps: silica encapsulation of the TDBC donor, preparation of the silane functionalized ATTO 590 acceptor, and subsequent covalent integration of the acceptor onto the silica surface of TDBC J-aggregates.

In the first step, TDBC molecules were assembled into two-dimensional J-aggregate sheets in aqueous solution in the presence of (3-aminopropyl)triethoxysilane (APTES) (Fig.1, Step 1). At approximately neutral pH, the terminal amine of APTES (pKa~9.6) is predominantly protonated (>99% at pH 7), generating positively charged ammonium groups that interact electrostatically with the negatively charged sulfobutyl groups of TDBC.[44,45] These positively charged amine groups of APTES interact electrostatically with the negatively charged sulfobutyl groups of TDBC, reducing repulsion between neighboring TDBC molecules and promoting close intermolecular packing and aggregation. Concurrently, hydrolysis of the ethoxysilane groups of APTES generates aminopropylsilanetriol which undergoes condensation to form an initial siloxane network (Si-O-Si) at the aggregate surface. This APTES rich interfacial layer subsequently provides nucleation sites for further silica growth following addition of tetraethyl orthosilicate (TEOS), resulting in a continuous silica coating around the TDBC sheets. Importantly, the silica layer physically separates the excitonic donor from species subsequently attached to its surface while retaining the optical characteristics of the underlying J-aggregate.[23]

The molecular acceptor was functionalized separately prior to coupling with the silica-encapsulated J-aggregate donor (Fig. 1, Step 2). ATTO 590 NHS ester was reacted with the primary amine of APTES in DMSO (anhydrous, ≥ 99.9%), forming an amide-linked ATTO 590-APTES conjugate while retaining the trialkoxysilane functionality required for subsequent surface coupling. Formation of the ATTO 590-APTES conjugate was confirmed by high resolution LC-MS (Supplementary, Figs. S1-S4) which showed a product ion at m/z = 794.4207, in excellent agreement with the calculated value of 794.4206 for the expected conjugate ($\Delta$ = 0.13 ppm).

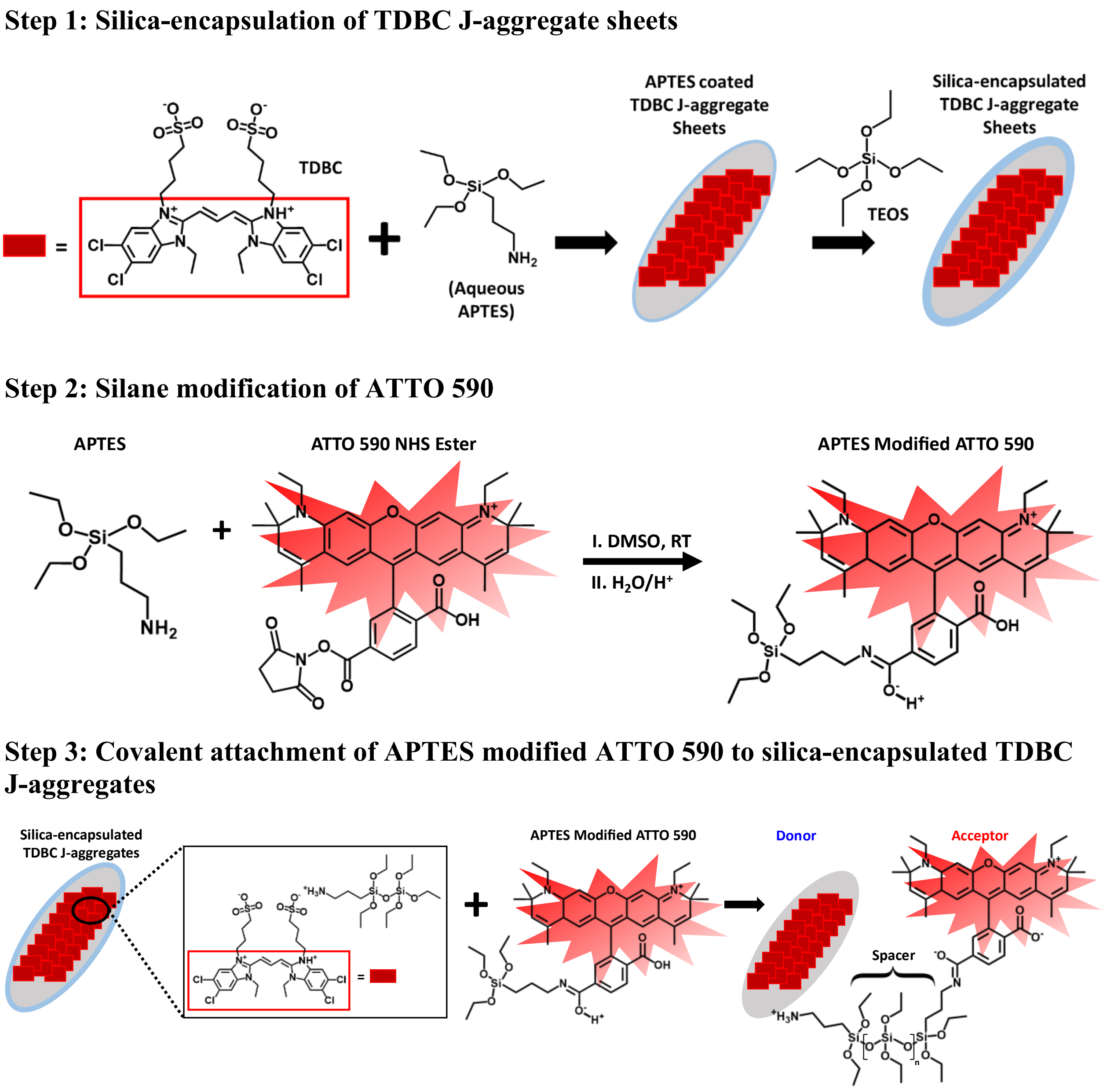

**Figure 1. Synthesis of silica-encapsulated TDBC J-aggregate-ATTO 590 donor-acceptor hybrids.** Step 1: TDBC molecules self-assemble into two-dimensional J-aggregate sheets in the presence of (3-aminopropyl)triethoxysilane (APTES), followed by silica growth through hydrolysis and condensation of APTES and tetraethyl orthosilicate (TEOS) to form silica-encapsulated TDBC donor J-aggregate assemblies. Step 2: ATTO 590 NHS ester is reacted with APTES in dry DMSO through amide formation to produce a silane-functionalized ATTO 590-APTES conjugate. Step 3: The ATTO 590-APTES conjugate is incorporated onto the silica-encapsulated TDBC J-aggregate surface through siloxane condensation, yielding a covalently integrated donor-acceptor architecture in which the silica interface spatially separates the two-dimensional excitonic donor from localized molecular acceptors.

In the final step, the silane-modified ATTO 590 was coupled to the silica-encapsulated TDBC J-aggregate sheets (Fig. 1, Step 3). Upon transfer to the aqueous reaction medium, the remaining ethoxysilane groups of the ATTO 590-APTES conjugate hydrolyze to form silanols, which subsequently undergo condensation with silanol groups present at the surface of the silica shell. This process incorporates the molecular acceptor into the outer silica network through covalent Si-O-Si linkages, yielding a hybrid structure in which the TDBC J-aggregate acts as the excitonic donor and ATTO 590 as the localized molecular acceptor. Following the coupling reaction, the constructs were purified using 100 kDa molecular-weight-cutoff centrifugal filter units to remove any unbound ATTO 590. The amount of ATTO 590 introduced during the third step was varied systematically to prepare a series of samples with different donor/acceptor ratios. Because the acceptor is coupled through the silane interface rather than adsorbed directly onto the TDBC assembly, its loading can be varied while preserving the excitonic structure of the J-aggregate. The resulting constructs provide a platform for studying energy transfer from an extended two-dimensional excitonic donor to surface-bound molecular acceptors with minimal perturbation of the supramolecular donor structure.

***Structural characterization and control of donor-acceptor separation***

Preservation of the two-dimensional donor morphology throughout silica encapsulation and acceptor functionalization was first evaluated by electron microscopy (Fig. 2a-c). Bare TDBC J-aggregates form elongated, sheet-like structures with lateral dimensions extending to the micron scale (Fig. 2a). Following silica encapsulation, the characteristic sheet morphology is retained (Fig. 2b), indicating that growth of the inorganic shell does not disrupt the underlying supramolecular architecture. The silica-encapsulated sheets also exhibit increased lateral dimensions, consistent with electrostatic screening by positively charged silanes. The resulting structures display aspect ratios of approximately 4. Importantly, the same overall morphology is preserved after covalent incorporation of ATTO 590 (Fig. 2c). These observations show that the silica layer provides a structurally protective and chemically addressable interface that enables acceptor functionalization without substantially altering the macroscopic morphology of J-aggregates.

The vertical dimensions of the silica-encapsulated TDBC sheets were further examined by atomic force microscopy (AFM) (Fig. 2d). Height profiles extracted across the regions indicated in Fig. 2d are shown in Fig. 2e and reveal an average thickness of ~8.9 nm for the silica-encapsulated TDBC J-aggregates. The thickness of bare TDBC J-aggregate sheets was previously measured to be ~2.7 nm.[23] Assuming symmetric silica growth on the two faces of the J-aggregate sheet, the average silica shell thickness is therefore estimated to be ~3.1 nm per side. In addition to the inorganic spacer, the molecular linker connecting ATTO 590 to the silica surface contributes to the effective donor-acceptor separation. The linker length was defined as the direct distance between the APTES silicon atom, representing the silica-side anchoring position, and the aromatic carbon at which the linker connects to the ATTO 590 chromophore. The geometry of the ATTO 590-APTES conjugate was modeled using Avogadro, yielding a linker length of approximately 7.67 Å (Fig. 2f). The effective donor-acceptor separation was consequently treated as the combined contribution from the silica layer and the organic linker, which is 3.9 nm.

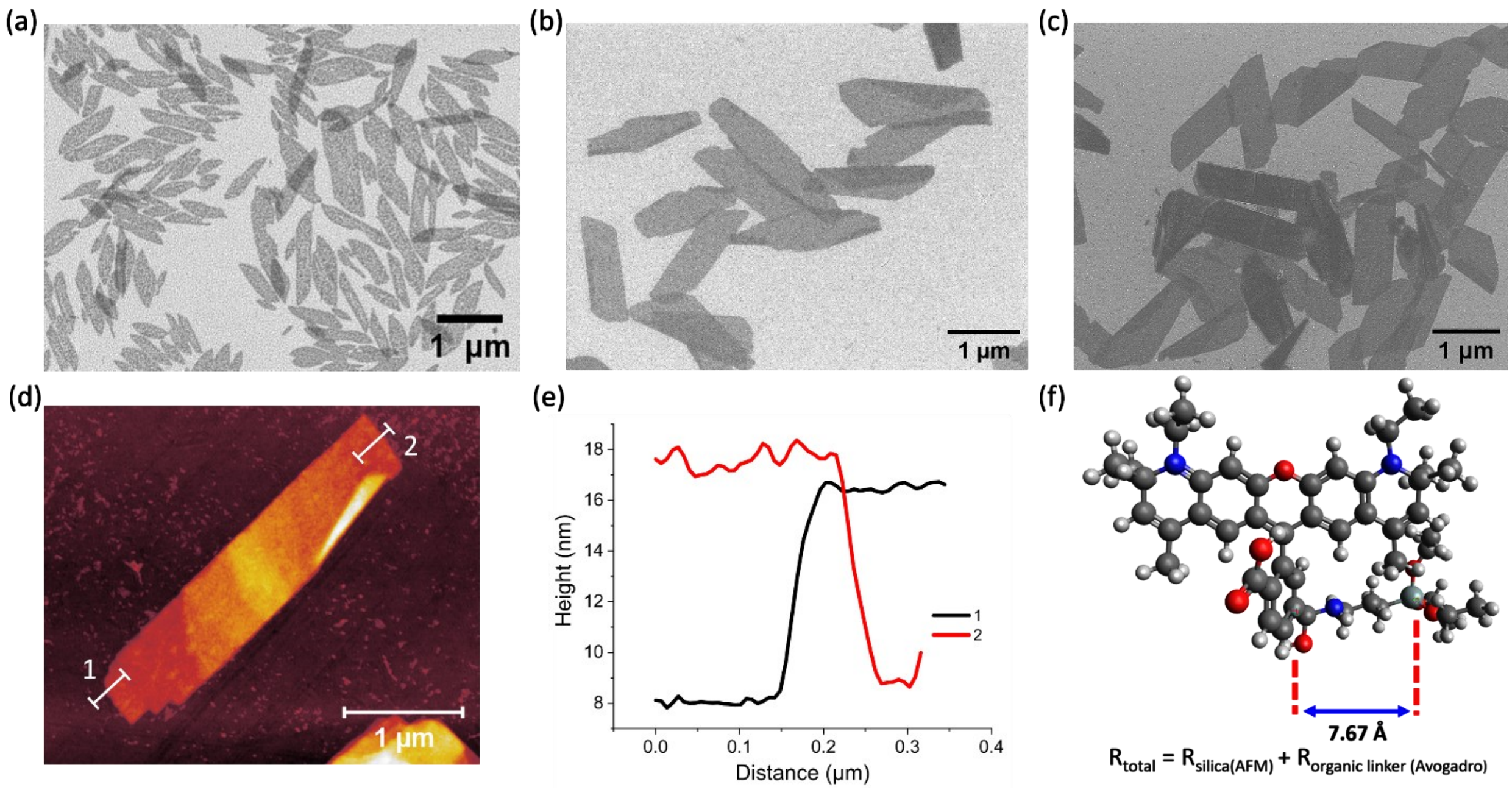

**Figure 2. Structural characterization and donor-acceptor separation.** SEM images of (a) bare TDBC J-aggregate sheets, (b) silica-encapsulated TDBC J-aggregate sheets, and (c) silica-encapsulated TDBC J-aggregate-ATTO 590 hybrids following covalent acceptor functionalization. (d) AFM image of an individual silica-encapsulated TDBC J-aggregate sheet. (e) Representative height profiles extracted across the two-line scans indicated in panel (d), showing the vertical dimensions. (f) Molecular model of the ATTO 590-APTES linker optimized in Avogadro, used to estimate the organic linker length contributing to the donor-acceptor separation. The total donor-acceptor separation was estimated from the combined silica-shell thickness and molecular linker length.

### *Steady-state spectroscopic evidence for excitonic energy transfer*

To quantify acceptor loading in the hybrid constructs, the composition was expressed in terms of TDBC monomer equivalents relative to ATTO 590. After formation of the silica-encapsulated TDBC-ATTO 590 hybrids, the constructs were isolated by centrifugation[46] and subsequently disassembled in methanol to recover the constituent chromophores for quantitative analysis. The concentrations of TDBC and ATTO 590 were determined independently by standard addition method, using the corresponding molecular standards (Supplementary, Fig. S5). The resulting values therefore represent the molar ratio of TDBC monomer equivalents to covalently incorporated ATTO 590 acceptors. Accordingly, the loading ratios reported in Fig.3, from 90:1 to 19:1, correspond to progressively increasing ATTO 590 within donor-acceptor hybrids.

The steady-state optical properties of the donor and acceptor establish the spectral conditions required for resonance energy transfer. Silica-encapsulated TDBC J-aggregates exhibit a characteristic narrow J-band centered at 586 nm, with nearly coincident absorption and emission spectra (Fig. 3a).[22,23] In contrast, APTES-functionalized ATTO 590 exhibits a broader absorption band centered at approximately 601 nm that substantially overlaps with the TDBC J-aggregate emission, along with a fluorescence band centered near 620 nm (Fig. 3b). The strong spectral overlap ($J = 3.57 \times 10^{15}$ $M^{-1}$ $cm^{-1}$ $nm^{4}$) provides the spectral basis for dipole-mediated energy transfer between the delocalized TDBC excitonic donor and the localized ATTO 590 acceptor (Supplementary, Fig. S6).[29,30]

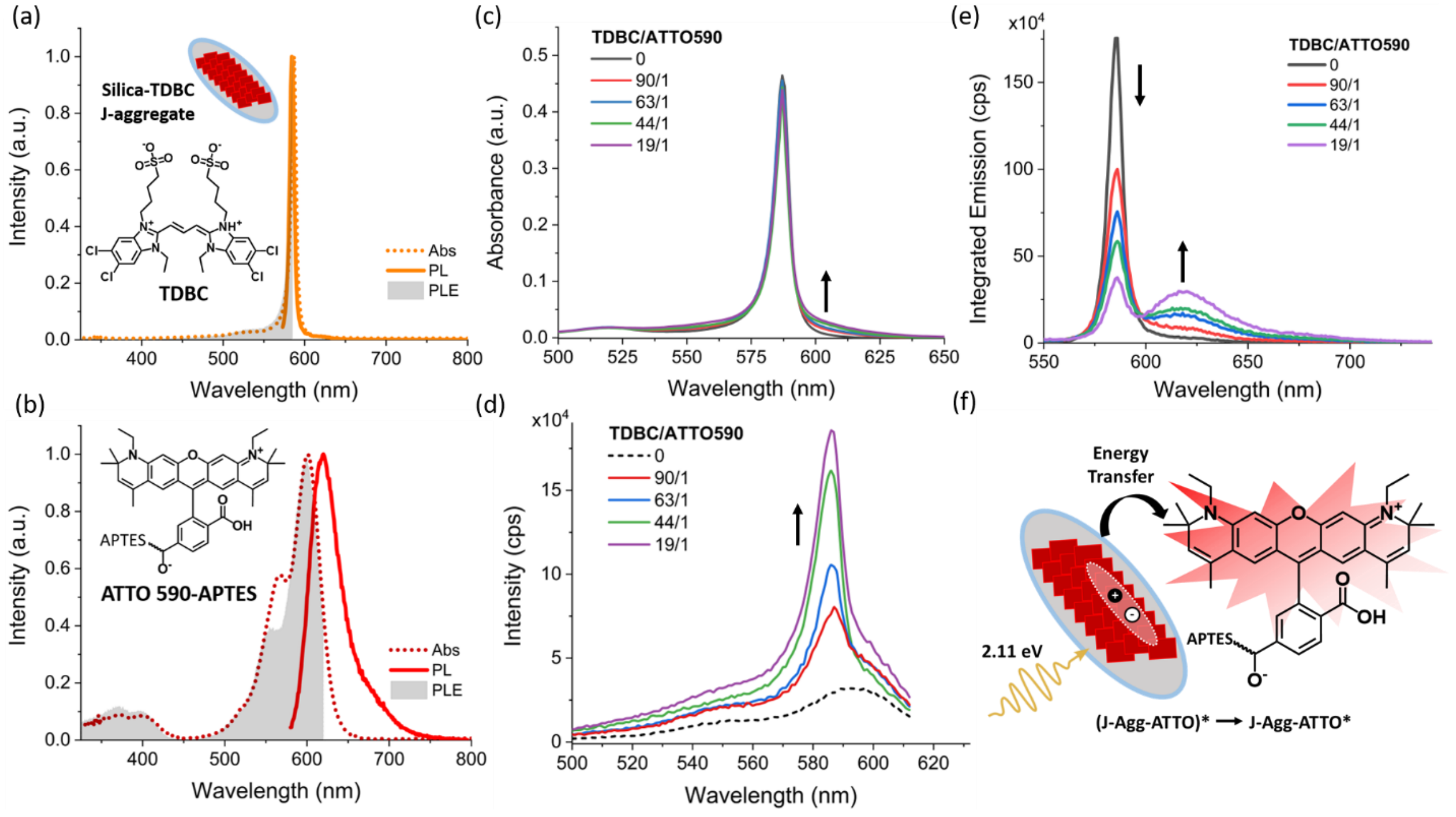

**Figure 3. Steady-state spectroscopic characterization of energy transfer in silica-encapsulated TDBC J-aggregate-ATTO 590 hybrids.** Normalized absorption (dotted line), photoluminescence (PL; solid line), and photoluminescence-excitation (PLE; shaded) spectra of (a) silica-encapsulated TDBC J-aggregates, showing the narrow J-band characteristic of the excitonic donor, and (b) ATTO 590-APTES, illustrating the substantial spectral overlap between donor emission and acceptor absorption. (c) Absorption spectra of silica-encapsulated TDBC-ATTO 590 hybrids with increasing acceptor loading at constant donor concentration. (d) PLE spectra of the hybrid series with emission monitored at 620 nm, showing an increasing contribution from excitation through the TDBC J-band with increasing ATTO 590 loading. (e) PL spectra of the hybrid series recorded at $\lambda_{exc}$= 450 nm, revealing progressive quenching of TDBC donor emission at 586 nm accompanied by enhanced ATTO 590 emission around 620 nm, consistent with acceptor sensitization. (f) Schematic representation of excitation of the two-dimensional TDBC J-aggregate donor followed by energy transfer to a localized ATTO 590 molecular acceptor. TDBC/ATTO 590 ratios denote the molar ratio of TDBC monomer equivalents to covalently incorporated ATTO 590 molecules; "0" denotes silica-encapsulated TDBC J-aggregates without ATTO 590.

Integration of increasing ATTO 590 concentration does not substantially alter the position or line shape of the TDBC J-band in the absorption spectra (Fig. 3c). The persistence of this feature across the hybrid series indicates that the collective excitonic behavior of the TDBC assembly is retained following acceptor functionalization. Increasing acceptor loading produces a growing long wavelength contribution to the absorption spectrum, consistent with increasing ATTO 590. The absence of a pronounced shift or broadening of the J-band is significant because it indicates that the spectroscopic changes observed in emission are not primarily driven by disruption of the underlying J-aggregate structure. Increasing ATTO 590 loading results in a progressive decrease in the TDBC donor emission at 586 nm, accompanied by a concomitant increase in emission near 620 nm, where ATTO 590 fluorescence dominates. The correlated loss of donor emission and growth of acceptor emission is consistent with an additional non-radiative relaxation pathway in which excitation initially associated with the TDBC J-aggregate is transferred to the molecular acceptor. The systematic evolution of the spectra with acceptor loading further argues against a simple static perturbation of the donor and instead supports increasing coupling between the two components as the number of surface bound acceptors is increased. Photoluminescence excitation (PLE) selectively monitors emission from the acceptor region around 620 nm while varying the excitation wavelength. With increasing ATTO 590 loading, a progressively stronger excitation feature develops near the TDBC J-band at 586 nm (Fig. 3d). This indicates that excitation absorbed by the J-aggregate increasingly contributes to emission detected in the ATTO 590 spectral region. This sensitized excitation pathway consists of energy being harvested by the delocalized TDBC excitonic manifold and subsequently transferred to localized excited electronic states of ATTO 590, covalently bound in close proximity to the J-aggregate surface as shown in the schematic: Fig. 3f. Taken together, the preservation of the J-band absorption, donor emission quenching, acceptor emission enhancement, and emergence of the donor excitation signature in the acceptor monitored PLE spectra provide mutually consistent steady state evidence for energy transfer within the silica-encapsulated TDBC-ATTO 590 hybrids.

***Energy transfer efficiencies from complementary steady state and time resolved measurements***

For the 19:1 TDBC/ATTO 590 construct, energy transfer efficiency was quantified using three independent observables: donor emission quenching, sensitized acceptor emission, and shortening of the donor excited-state lifetime. The corresponding equations, fitting procedures, and numerical calculations are provided in the Supplementary Information (Section 6). The donor quenching analysis was performed by resolving the TDBC emission from the broader ATTO 590 contribution in the hybrid spectrum (Fig. 4a). The isolated donor feature remains centered near 586 nm and retains a linewidth comparable to that of the donor only sample, with full widths at half maximum of approximately 8 and 9 nm, respectively. Donor-quenching analysis gave an energy transfer efficiency of 79% from the peak emission intensity and 75% from the integrated donor emission area.

In Fig 4b (and Supplementary, Fig. S9), the emission spectrum of TDBC-ATTO 590 conjugate contains overlapping contributions from residual TDBC emission, direct excitation of ATTO 590, and emission arising from donor-mediated sensitization. After correcting for residual donor emission and direct ATTO 590 excitation, the sensitized acceptor emission yielded an energy transfer efficiency of approximately 76%. The fluorescence quantum yields of silica-encapsulated TDBC J-aggregates ($\Phi_D = 0.94$) and ATTO 590-APTES ($\Phi_A = 0.70$) were determined using the relative method with Rhodamine 6G and Rhodamine 101 as reference standards, respectively (Supplementary, Figs. S7, S8).[47,48] Correcting the sensitized and residual donor emission for their respective photoluminescence quantum yields increased the sensitization-derived efficiency to approximately 81%.The close agreement between donor quenching and acceptor sensitization measurements provides complementary steady state evidence that a substantial fraction of the excitation initially associated with the TDBC J-aggregate donor is redirected into ATTO 590 emission.

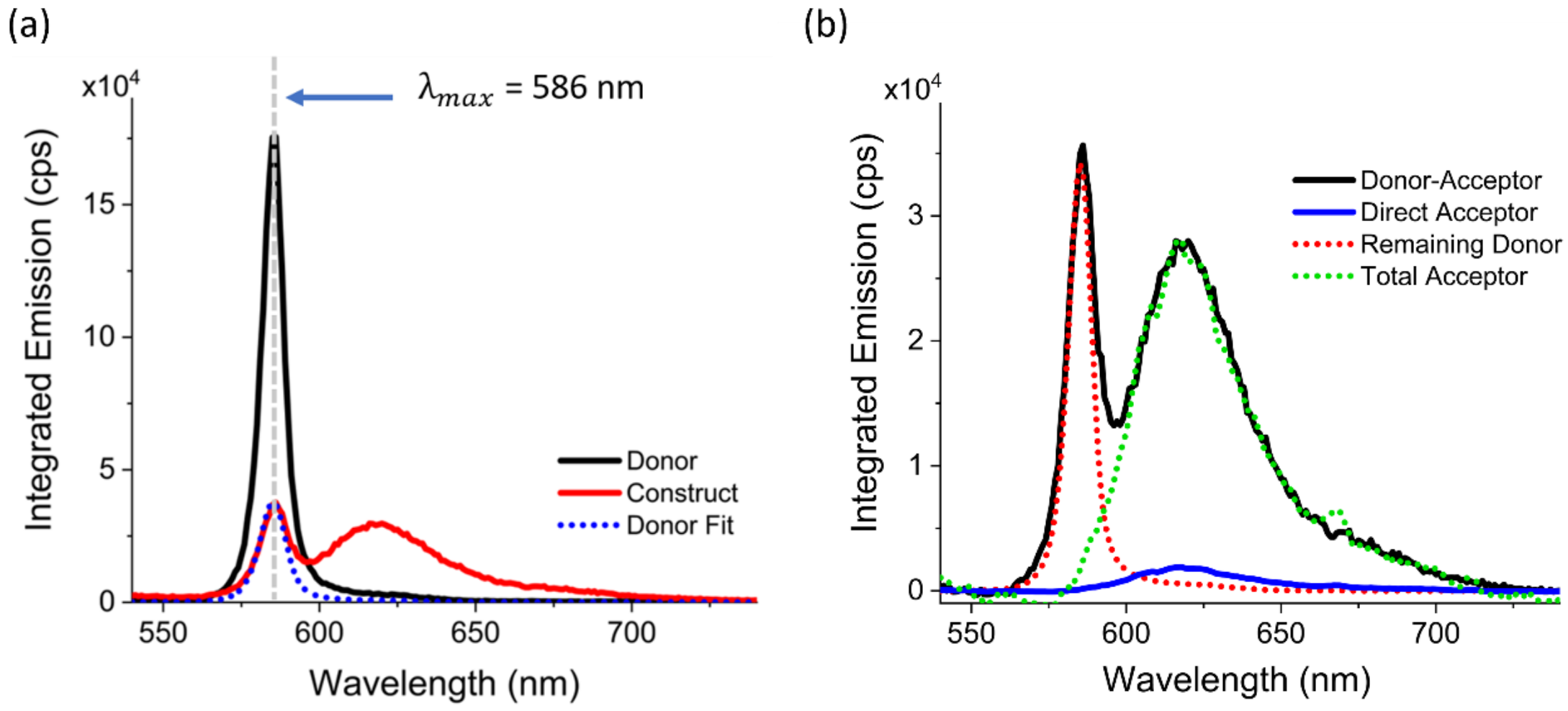

**Figure 4. Quantification of steady-state energy transfer efficiency in the 19:1 TDBC/ATTO 590 donor-acceptor construct.** (a) Comparison of the donor-only and donor-acceptor emission spectra. Donor quenching was quantified from the emission intensity at the TDBC peak (586 nm) and from the integrated area of the fitted residual donor contribution. (b) Spectral decomposition of the donor-acceptor construct into the remaining TDBC donor emission and total ATTO 590 emission. The directly excited acceptor contribution, determined from the acceptor-only control, was separated from the total acceptor emission to obtain the sensitized ATTO 590 contribution used to quantify energy transfer.

Time resolved measurements provide a direct measure of the additional excited state decay pathway introduced by the acceptor (Supplementary, Fig. S10). The silica-encapsulated TDBC donor exhibits a fitted emission lifetime of ~207 ps, whereas incorporation of ATTO 590 shortens the donor decay to ~93 ps. This lifetime reduction is consistent with the opening of an additional non-radiative decay channel associated with energy transfer to the molecular acceptor. The lifetime-derived transfer efficiency was 55%, lower than the efficiencies obtained from steady-state donor quenching and acceptor sensitization (~75-81%). The difference likely reflects heterogeneity within the donor-acceptor ensemble. Strongly coupled donors can contribute substantially to steady-state donor quenching and sensitized acceptor emission while contributing relatively little to the measured donor decay because their fluorescence is rapidly depleted. The time-resolved measurement is therefore weighed more strongly toward the remaining emissive donor population, which can lead to a lower apparent transfer efficiency. Therefore, the lifetime derived value is treated as an effective ensemble-averaged transfer efficiency rather than being expected to quantitatively reproduce the steady state estimates.

### *Plane-to-point Förster coupling in a two-dimensional excitonic donor*

To describe energy transfer from the extended TDBC J-aggregate donor to localized ATTO 590 acceptors, we developed a generalized Förster treatment based on the distributed dipole model that accounts for donor coherence, dipole orientation, and the two-dimensional donor geometry (Supplementary, Figs. S11-S14). The experimental system[23] provides a critical test of the Förster theory and its generalization. The rate calculation is based on the distributed dipole model,[49] which was initially proposed as a numerical approach and has been adopted as the basis for the scaling analysis of interwire FRET[23] and collective energy transfer in optical cavities.[50] The analysis reported in this paper predicts different scaling regimes as a function of z, the donor plane-to-acceptor molecule distance, as illustrated in Fig. 5:

(1) $z > r_{coh}, r_{agg}$ $1/z^6$
(2) $r_{agg} > z > r_{coh}$ $1/z^4$
(3) $r_{agg} > r_{coh} > z$ $1/z^4$ (with reduced numerator)
(4) $r_{coh} \sim r_{agg} > z$ $1/z^4$ (the numerator approaching zero as coherent size increases)

where $r_{coh}$ is the coherence size (i.e., localization length) of the TDBC J-aggregate and $r_{agg}$ is the physical size of the J-aggregate. The prediction holds strictly for the in-plane ordered dipole array and reveals a complex distance dependence with a possible turnover in between near-field and far-field energy transfer.

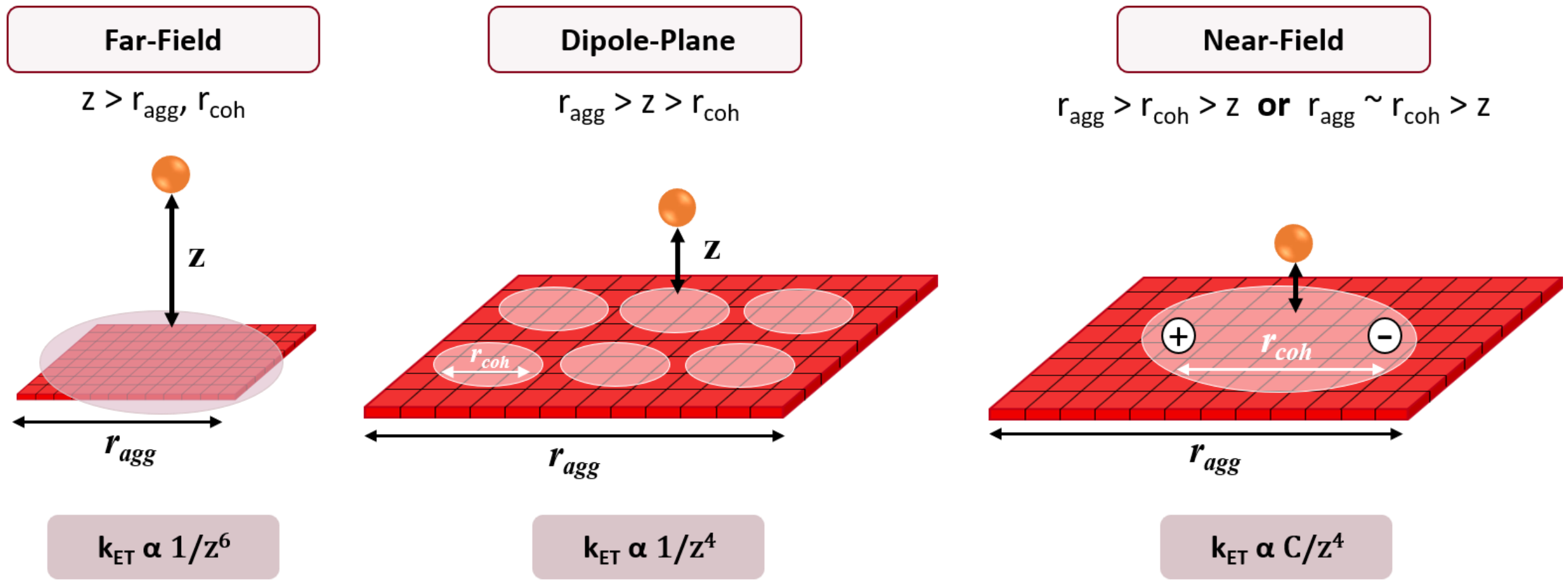

**Figure 5. Schematic illustration of the distance-dependent energy-transfer regimes in the generalized Förster model.** In the far-field regime, where the donor-acceptor separation (z) exceeds the physical aggregate size $r_{agg}$, the two-dimensional aggregate behaves as an effective point dipole and $k_{ET} \propto z^{-6}$. For $r_{agg} > z > r_{coh}$, where $r_{coh}$ is the exciton coherence length, integration of the Förster interaction over the two-dimensional donor plane gives the dipole-plane scaling $k_{ET} \propto z^{-4}$. In the near-field regime, $z < r_{coh}$, the acceptor resolves the spatial distribution of transition dipoles within a coherent donor domain, leading to partial cancellation of their contributions and a reduced coupling coefficient while retaining the $z^{-4}$ distance dependence, such that $k_{ET} \propto z^{-4}$, where (C) depends on the coherence size.

Regime (1) corresponds to the far-field coupling described by the point-dipole interaction, where the effective dipole of the 2D aggregate is proportional to the number of molecules $N_{agg}$. Regime (2) corresponds to the dipole-plane coupling, which is essentially the integration of the Förster over the 2D plane. Regime (4) corresponds to the fully coherent near-field coupling, where the collective dipole array on the 2D plane leads to reduced terminal charges on the order of $1/\sqrt{N_{coh}}$. Then the effective charge-dipole coupling scales as $1/z^2\sqrt{N_{coh}}$ and the resulting Förster rate scales as $1/z^4 N_{coh}$, which vanishes for infinitely large coherent size. As a result, this regime has the same scaling as regime (2) but with different coefficients. Regime (3) is a mixed case, where the coherent cluster directly opposite to the acceptor is in the near-field but all the other coherent clusters are in the far-field. Since both near- and far-field couplings share the same scaling, the net effect is the reduced coefficient that depends on the coherent size $N_{coh}$. Interestingly, this coefficient has a non-monotonic dependence on $N_{coh}$, which can exhibit a turnover as the coherent size is tuned by disorder and/or temperature.

The quantitative evaluation of the Forster rate involves not only z-dependence but also the spectral overlap.[49] The scaling relationship applies to both forward and backward rates, whereas the spectral overlap differentiates the two rates and imposes the detailed balance condition.[50] Incorporating the experimental spectral overlap, donor photophysical properties, acceptor surface density, and a donor-acceptor separation of 3.9 nm gives $k_{ET} = 4.77 \times 10^{10}$ $s^{-1}$, approximately 10 times faster than the intrinsic donor decay rate ($k_D = 4.83 \times 10^9$ $s^{-1}$), corresponding to a theoretical energy-transfer efficiency of approximately 91% for the 19:1 TDBC/ATTO 590 construct. This value represents an upper theoretical estimate because the quantitative calculation uses the idealized dipole-plane coupling expression and does not explicitly account for the coherence-dependent reduction in the coupling coefficient expected when the donor-acceptor separation becomes comparable to or smaller than the exciton coherence size. In this near-field regime, destructive interference within the coherent TDBC dipole array reduces the effective donor-acceptor coupling, leading to lower transfer efficiencies than predicted by the idealized calculation. A full derivation of the generalized Förster model is provided in Supplementary Information.

## CONCLUSION

We have established a hybrid excitonic architecture in which two-dimensional TDBC J-aggregate sheets act as extended light-harvesting donors and covalently integrated ATTO 590 molecules serve as localized acceptors. Silica encapsulation plays a central role in this design by preserving the collective excitonic properties and sheet-like morphology of the J-aggregate while providing a chemically addressable interface for controlled acceptor loading and donor-acceptor separation. Steady-state spectroscopy reveals substantial donor quenching and sensitized ATTO 590 emission, with transfer efficiencies approaching 81%, while time resolved measurements independently confirm the opening of an additional donor decay pathway following acceptor incorporation. To describe energy transfer across this hybrid, we developed a plane-to-point Förster treatment in which finite exciton coherence and the two-dimensional donor geometry led to an inverse fourth-power dependence of the transfer rate on donor-acceptor separation, distinct from conventional point-to-point Förster transfer. Together, these results establish silica-encapsulated J-aggregates as modular building blocks for hybrid excitonic materials, enabling delocalized excitation to be captured, transported and redistributed into localized molecular states.

## METHODS

### Synthesis of silica-encapsulated TDBC J-aggregates

J-aggregate samples were prepared using TDBC (FEW Chemicals, S0046) monomers purified by the previously reported methanol-washing procedure.[23] Silica encapsulation was carried out inside a nitrogen glove bag. Immediately before each preparation, APTES (99%, Sigma Aldrich, 919-30-2) and TEOS (99%, Sigma Aldrich, 78-10-4) were freshly diluted in methanol (>99.9%, VWR, EM-MX0488-1) to prepare 1:249 v/v solutions. In a cleaned vial, 10 μL of freshly diluted APTES was added to 612 μL of ultrapure water (Milli-Q, 18 MΩ·cm) and mixed by shaking for several minutes. Subsequently, 88 μL of a 1.29 mM TDBC monomer stock solution in methanol was added and mixed, followed by the addition of 10 μL of freshly diluted TEOS. The resulting solution was mechanically shaken and vortexed at maximum speed several times. The vial was then wrapped in aluminum foil and stored in the dark overnight to allow silica condensation to proceed. Freshly prepared silica-encapsulated TDBC J-aggregates were used for measurements or for preparation of donor-acceptor constructs on the following day.

### Silane modification of ATTO 590

A 0.274 mM stock solution of ATTO 590 NHS ester was prepared by dissolving 1 mg of ATTO 590 NHS ester in 4.630 mL of anhydrous DMSO. The stock solution was stored in an airtight container at -20 °C, protected from moisture and light until use. For silane modification, APTES was freshly diluted in anhydrous DMSO (1:149 v/v, APTES:DMSO), and 10 μL of the diluted APTES solution was added to 200 μL of the ATTO 590 NHS ester stock solution and thoroughly mixed. The mixture was allowed to react for 30 min to facilitate amide bond formation between the NHS ester group of ATTO 590 and the primary amine of APTES. APTES was used in controlled molar excess (ATTO 590:APTES = 1:5) to promote efficient conjugation while limiting the amount of unreacted APTES, as excessive APTES can induce agglomeration of the silica-encapsulated TDBC J-aggregates during subsequent functionalization. Following the reaction, 210 μL of ultrapure water was added immediately before introducing the conjugate into the silica-encapsulated TDBC J-aggregate solution to promote hydrolysis of the ethoxysilane groups and facilitate subsequent coupling to the silica surface. The reaction mixture was used without further purification. All donor-acceptor constructs were prepared using freshly synthesized ATTO 590-APTES conjugates.

### LC-MS characterization of the ATTO 590-APTES conjugate

For LC-MS characterization, a 0.274 mM solution of ATTO 590 NHS ester was prepared separately in acetonitrile (Supergradient HPLC Grade, 99.9+%, 75-05-8) instead of DMSO to ensure compatibility with LC-MS analysis. APTES was freshly diluted in acetonitrile (1:149 v/v, APTES:acetonitrile), and 10 μL of the diluted APTES solution was added to 200 μL of the ATTO 590 NHS ester solution. The mixture was thoroughly mixed and allowed to react for 30 min to facilitate amide bond formation between the NHS ester group of ATTO 590 and the primary amine of APTES. No water was intentionally introduced during conjugation or subsequent sample preparation to minimize premature silane hydrolysis and condensation. Prior to LC-MS analysis, the reaction mixture was further diluted with acetonitrile to an approximate analyte concentration of 100 ppm (0.1 mg/mL). Solutions of the individual ATTO 590 NHS ester and APTES starting materials in acetonitrile (0.1 mg/mL) were also analyzed as controls to aid the assignment of chromatographic peaks and mass spectra.

LC-MS measurements were performed using an Agilent 6545 Q-TOF high-resolution mass spectrometer coupled to an Agilent Infinity 1260 LC system equipped with a C18 reversed-phase column. The instrument was operated using a Jet Stream electrospray ionization (ESI) source in positive ion mode. Water containing 0.1% formic acid and acetonitrile containing 0.1% formic acid were used as mobile phases A and B, respectively. The gradient was held at 95% A/5% B for 2 min, followed by a linear increase to 100% B over 13 min. The mobile phase was maintained at 100% B for 2 min, returned to 95% A/5% B at 18 min, and followed by a 2 min post-run re-equilibration. The column temperature was maintained at 25 °C, the flow rate was 0.3 mL $min^{-1}$, and the injection volume was 1 μL. Mass spectra were acquired over an m/z range of 100-900 and processed using Agilent MassHunter software. Molecular species were assigned by comparing the experimentally observed accurate m/z values with the calculated exact masses of the expected ATTO 590-APTES amide conjugate and its partially and fully hydrolyzed silane derivatives. Mass assignments accounted for the intrinsically charged nature of ATTO 590, with molecular cations $[M]^+$.

**Synthesis of silica-encapsulated TDBC J-aggregate-ATTO 590 donor-acceptor hybrids**

A series of donor-acceptor hybrids with varying acceptor loading was prepared by adding 10.0-35.0 μL of freshly synthesized ATTO 590-APTES conjugate solution to 700 μL of freshly prepared silica-encapsulated TDBC J-aggregate dispersion that had been aged overnight to allow silica condensation to proceed. To maintain a consistent solvent environment across the series, a 50:50 (v/v) DMSO/water mixture was added to adjust the total volume of each sample to 800 μL, resulting in a final nominal DMSO concentration of approximately 6.25% (v/v). The mixtures were mechanically shaken and vortexed to facilitate uniform dispersion and interaction of the silane-functionalized acceptors with the silica surface. The samples were then wrapped in aluminum foil and incubated overnight at room temperature in the dark, without further agitation, to promote coupling of the ATTO 590-APTES conjugates to the silica surface. Following overnight incubation, the donor-acceptor hybrids were purified using 100 kDa molecular-weight-cutoff centrifugal filter units. Prior to filtration, the filter units were preconditioned with an aqueous solution containing approximately 6.25% (v/v) DMSO to equilibrate the membranes and minimize nonspecific adsorption and variability in sample recovery during filtration. The samples were then centrifuged at 8000 rpm for 10 min (acceleration setting = 9) to remove unbound ATTO 590-APTES and other low-molecular-weight species. The purified donor-acceptor hybrids were redispersed in 800 μL of ultrapure water prior to optical characterization. The resulting donor-acceptor hybrids exhibited experimentally determined TDBC/ATTO 590 molar ratios of approximately 90:1, 63:1, 44:1, and 19:1, as quantified by standard addition following separation and disassembly of the constructs (Supplementary, Fig. S5). These experimentally determined molar ratios were assumed to be representative of the corresponding 100 kDa-filtered donor-acceptor hybrids used for optical characterization.

**Determination of TDBC/ATTO 590 molar ratios**

The TDBC and ATTO 590 contents were independently determined using the standard-addition method. For TDBC quantification, a donor-only reference was prepared using 700 μL of the same silica-encapsulated TDBC J-aggregate dispersion used for the donor-acceptor hybrids. The ATTO 590-APTES conjugate was omitted, and the total volume was adjusted to 800 μL using a 50:50 (v/v) DMSO/water mixture to maintain nominally equivalent solvent conditions. The donor-only reference underwent the same overnight incubation as the donor-acceptor hybrids. Following incubation, the donor-only reference and donor-acceptor hybrid dispersions were centrifuged at 8000 rpm for 25 min (acceleration setting = 9) to sediment the silica-encapsulated structures and separate them from soluble, unbound species. The supernatants were carefully removed, and the resulting pellets were air-dried to remove residual solvent. The pellets were then redispersed in 1 mL of methanol and vigorously vortexed to facilitate disassembly of the J-aggregates. The samples were subsequently heated at 70 °C for 5-10 min in capped vials to promote dissolution while minimizing solvent evaporation and allowed to cool to room temperature before analysis.

For the standard-addition measurements, the resulting methanol solutions were divided into 200 μL aliquots. TDBC content was determined by adding 10-40 μL of a 1.29 mM TDBC monomer standard solution in methanol to aliquots of the donor-only reference. ATTO 590 content was determined by adding 5-20 μL of a 0.274 mM ATTO 590 NHS ester standard solution in anhydrous DMSO to aliquots of the donor-acceptor hybrid samples. Additional methanol or DMSO was added as appropriate to adjust each sample to a final volume of 240 μL while maintaining a constant solvent composition within each standard-addition series. Absorbance measurements were performed for each sample, and linear fits of absorbance as a function of added chromophore concentration were used to determine the native TDBC and ATTO 590 concentrations from the magnitudes of their respective x-intercepts. The TDBC concentration obtained from the identically processed donor-only reference was used to estimate the TDBC content of the donor-acceptor hybrids. The corresponding TDBC/ATTO 590 molar ratios were calculated from these concentrations, yielding approximately 90:1, 63:1, 44:1, and 19:1 (Supplementary, Fig. S5).

**AFM measurements**

Atomic force microscopy (AFM) measurements were performed in air using a Cypher VRS1250 microscope (Asylum Research, Oxford Instruments) operated in tapping mode. A fast-scanning, high-frequency silicon cantilever (FS1500AUD, Asylum Research, Oxford Instruments; nominal resonance frequency of 1.5 MHz and spring constant of 6 $Nm^{-1}$) was used for imaging. Samples were prepared by spin-coating J-aggregate dispersions onto glass substrates hydrophilized by oxygen plasma treatment. For each deposition cycle, a 5 µL aliquot of the J-aggregate dispersion was deposited onto a stationary substrate, followed by spin-coating at 1200 rpm for 2 min. The deposition procedure was repeated as needed until sufficient surface coverage was achieved. AFM images were processed using Gwyddion software, and height measurements were obtained from representative line profiles across individual J-aggregate structures.

**SEM measurements**

Bare TDBC J-aggregates, silica-encapsulated TDBC J-aggregates, and silica-encapsulated TDBC J-aggregate-ATTO 590 donor-acceptor hybrids were deposited onto $Si/SiO_2$ substrates by spin-coating. Prior to deposition, the substrates were rendered hydrophilic by oxygen plasma treatment. A 5 µL aliquot of each sample dispersion was deposited onto the substrate and spin-coated at 800 rpm for 2 min. The samples were subsequently allowed to dry in air before imaging. Scanning electron microscopy (SEM) images were acquired using a ZEISS SIGMA HD VP scanning electron microscope operated at an accelerating voltage of 1.0 kV, with a working distance of approximately 4.8 mm and a 30 µm objective aperture. Secondary-electron images were collected using the InLens detector. All samples were imaged without additional conductive coating.

**Absorption and fluorescence spectroscopy**

UV-vis absorption and steady-state photoluminescence spectra were recorded using a Cary 5000 UV-vis-NIR spectrophotometer (Agilent Technologies) and a FluoroMax spectrofluorometer (HORIBA Scientific), respectively. Measurements were performed using demountable quartz cuvettes (Starna Cells) with an optical path length of 0.1 mm. Approximately 60 µL of each sample dispersion was introduced into the cuvette, and excess solution was carefully wiped from the external surfaces after assembly. The same cuvette configuration was used for both absorption and photoluminescence measurements. The excitation wavelengths used for fluorescence measurements are specified in the corresponding figure captions.

**Quantum yield (QY)**

Photoluminescence quantum yields (PLQYs) were determined using a relative method.[23] Rhodamine 6G in methanol (Sigma, 252433-250MG; $\Phi = 0.93$) and rhodamine 101 in ethanol ($\Phi = 0.91$) were used as reference standards for the silica-encapsulated TDBC J-aggregates and silane-modified ATTO 590, respectively.[47,48] Concentration series of each reference were prepared by dilution in the corresponding solvent. Similarly, dilution series of silica-encapsulated TDBC J-aggregates and silane-modified ATTO 590 were prepared from their respective stock solutions using water and a 50:50 (v/v) DMSO/water mixture, respectively, and allowed to equilibrate for 3 h before measurement. Absorption and photoluminescence spectra were collected using demountable quartz cuvettes with an optical path length of 0.1 mm. Rhodamine 6G and the silica-encapsulated TDBC J-aggregates were excited at 500 and 565 nm, respectively, whereas rhodamine 101 and silane-modified ATTO 590 were excited at 550 and 580 nm, respectively. For each concentration series, the integrated photoluminescence intensity was plotted as a function of absorptance, and the slope was determined by linear regression. PLQYs were calculated relative to the corresponding reference standards using the fitted slopes, with refractive indices of the respective solvents, and the excitation powers at each wavelength.

**Lifetime measurements**

Time-resolved photoluminescence measurements were performed using 515 nm pulsed laser excitation (PicoQuant Prima) at an excitation power of 180 nW and a repetition rate of 10 MHz. A 90/10 beamsplitter (Thorlabs, BS028) and a variable neutral-density filter were used to monitor and adjust the excitation power. The laser beam was focused onto the sample using a 20× objective (Olympus, PLN20×), producing an excitation spot with a full width at half maximum (FWHM) of approximately 45 μm. Photoluminescence was collected through a dichroic mirror (DMLP550) and imaged onto the entrance slit of a spectrometer (Acton SP2300). The emitted light was dispersed using a 150 grooves/mm grating centered at 550 nm. Streak images were acquired in photon-counting mode (Hamamatsu C10627) by accumulating 200,000 exposures of 10 ms each. Donor photoluminescence lifetimes were determined by fitting the measured decay profiles with a single-exponential decay function reconvolved with the measured temporal response to the 515 nm laser pulse. For donor-acceptor constructs, the TDBC emission contribution at each time point was extracted through global spectral fitting of the streak images to separate the overlapping donor and acceptor emission. Uncertainties in the extracted lifetimes were estimated from the local curvature of the nonlinear least-squares fitting function, using the Jacobian matrix evaluated at the optimized fitting parameters.

## ASSOCIATED CONTENT

**Supplementary Information Available:**

LC-MS characterization of the ATTO 590-APTES conjugate, Determination of TDBC/ATTO 590 molar ratios, Spectral overlap between TDBC J-aggregates and ATTO 590, Photoluminescence quantum yield measurements of silica-encapsulated TDBC J-aggregates, Photoluminescence quantum yield measurements of ATTO 590-APTES, Energy transfer efficiency calculations, Derivation of the plane-to-point Förster resonance energy transfer distance dependence.

## AUTHOR CONTRIBUTIONS

D. H. T. A. conceived and developed the project, designed the experiments, developed the silica-encapsulation and donor-acceptor coupling procedures, prepared all samples, performed the steady-state optical, structural, microscopic, LC-MS, quantum-yield and compositional measurements, and analyzed the resulting data. O. M. N. developed the instrumentation for time-resolved J-aggregate spectroscopy and performed and analyzed the lifetime measurements. J. C. developed the generalized Förster resonance energy-transfer framework, including the derivation of its distance dependence. D. H. T. A. extended the model by incorporating experimental spectral overlap and system-specific parameters to calculate the theoretical energy-transfer rate and efficiency. D. H. T. A. wrote the manuscript and Supplementary Information, with input and feedback from all authors. M. G. B. directed and supervised the project.

Notes: The authors declare no competing financial interest.

## ACKNOWLEDGMENTS

D. H. T. A. acknowledges support from the National Science Foundation (award CHE-2404354) and O. M. N. was supported by the US Department of Energy, Office of Science, Office of Basic Energy Sciences, Materials Chemistry Program under award number DE-FG02-07ER46454. J. C. was supported by DMREF (NSF-2324300), ACS-PRF, and MIT RSC fund. We thank the MIT.nano facility and the MIT Department of Chemistry Instrumentation Facility (DCIF) for assistance with AFM, SEM imaging, and LC-MS characterization.

## REFERENCES

1. Romero, E., Novoderezhkin, V. I. & van Grondelle, R. Quantum design of photosynthesis for bio-inspired solar-energy conversion. *Nature* **543**, 355-365 (2017).

2. Scholes, G. D., Fleming, G. R., Olaya-Castro, A. & van Grondelle, R. Lessons from nature about solar light harvesting. *Nat. Chem.* **3**, 763-774 (2011).

3. Mirkovic, T. *et al.* Light absorption and energy transfer in the antenna complexes of photosynthetic organisms. *Chem. Rev.* **117**, 249-293 (2017).

4. Scholes, G. D. & Rumbles, G. Excitons in nanoscale systems. *Nat. Mater.* **5**, 683-696 (2006).

5. Saikin, S. K., Eisfeld, A., Valleau, S. & Aspuru-Guzik, A. Photonics meets excitonics: natural and artificial molecular aggregates. *Nanophotonics* **2**, 21-38 (2013).

6. Sneyd, A. J. *et al.* Efficient energy transport in an organic semiconductor mediated by transient exciton delocalization. *Sci. Adv.* **7**, eabh4232 (2021).

7. Spano, F. C. The spectral signatures of Frenkel polarons in H- and J-aggregates. *Acc. Chem. Res.* **43**, 429-439 (2010).

8. Knoester, J. Modeling the optical properties of excitons in linear and tubular J-aggregates. *Int. J. Photoenergy* **2006**, 61364 (2006).

9. Fidder, H., Knoester, J. & Wiersma, D. A. Superradiant emission and optical dephasing in J-aggregates. *Chem. Phys. Lett.* **171**, 529-536 (1990).

10. Eisfeld, A. & Briggs, J. S. The J- and H-bands of organic dye aggregates. *Chem. Phys.* **324**, 376-384 (2006).

11. Eisele, D. M. *et al.* Utilizing redox-chemistry to elucidate the nature of exciton transitions in supramolecular dye nanotubes. *Nat. Chem.* **4**, 655-662 (2012).

12. Eisele, D. M., Knoester, J., Kirstein, S., Rabe, J. P. & Vanden Bout, D. A. Uniform exciton fluorescence from individual molecular nanotubes immobilized on solid substrates. *Nat. Nanotechnol.* **4**, 658-663 (2009).

13. Caram, J. R. *et al.* Room-temperature micron-scale exciton migration in a stabilized emissive molecular aggregate. *Nano Lett.* **16**, 6808-6815 (2016).

14. Kasha, M., Rawls, H. R. & El-Bayoumi, M. A. The exciton model in molecular spectroscopy. *Pure Appl. Chem.* **11**, 371-392 (1965).

15. Würthner, F., Kaiser, T. E. & Saha-Möller, C. R. J-aggregates: from serendipitous discovery to supramolecular engineering of functional dye materials. *Angew. Chem. Int. Ed.* **50**, 3376-3410 (2011).

16. Deshmukh, A. P. *et al.* Bridging the gap between H- and J-aggregates: classification and supramolecular tunability for excitonic band structures in two-dimensional molecular aggregates. *Chem. Phys. Rev.* **3**, 021401 (2022).

17. Spano, F. C. & Silva, C. H- and J-aggregate behavior in polymeric semiconductors. *Annu. Rev. Phys. Chem.* **65**, 477-500 (2014).

18. Didraga, C. *et al.* Structure, spectroscopy, and microscopic model of tubular carbocyanine dye aggregates. *J. Phys. Chem. B* **108**, 14976-14985 (2004).

19. Clark, J., Silva, C., Friend, R. H. & Spano, F. C. Role of intermolecular coupling in the photophysics of disordered organic semiconductors: aggregate emission in regioregular polythiophene. *Phys. Rev. Lett.* **98**, 206406 (2007).

20. Bondarenko, A. S., Jansen, T. L. C. & Knoester, J. Exciton localization in tubular molecular aggregates: size effects and optical response. *J. Chem. Phys.* **152**, 194302 (2020).

21. Deshmukh, A. P. *et al.* Near-atomic-resolution structure of J-aggregated helical light-harvesting nanotubes. *Nat. Chem.* **16**, 800-808 (2024).

22. Barotov, U. *et al.* Near-unity superradiant emission from delocalized Frenkel excitons in a two-dimensional supramolecular assembly. *Adv. Opt. Mater.* **11**, 2201471 (2023).

23. Thanippuli Arachchi, D. H. *et al.* Bright and fast emission from robust supramolecular J-aggregate nanostructures through silica-encapsulation. *ACS Nano* **18**, 20422-20434 (2024).

24. Gurunatha, K. L. & Dujardin, E. Tuning the optical coupling between molecular dyes and metal nanoparticles by the templated silica mineralization of J-aggregates. *J. Phys. Chem. C* **117**, 3489-3496 (2013).

25. Herman, K. *et al.* Individual tubular J-aggregates stabilized and stiffened by silica encapsulation. *Colloid Polym. Sci.* **298**, 937-950 (2020).

26. von Berlepsch, H. *et al.* Stabilization of individual tubular J-aggregates by poly(vinyl alcohol). *J. Phys. Chem. B* **107**, 14176-14184 (2003).

27. Anantharaman, S. B. *et al.* Enhanced room-temperature photoluminescence quantum yield in morphology controlled J-aggregates. *Adv. Sci.* **8**, 1903080 (2021).

28. Eisele, D. M. *et al.* Robust excitons inhabit soft supramolecular nanotubes. *Proc. Natl Acad. Sci. USA* **111**, E3367-E3375 (2014).

29. Stryer, L. & Haugland, R. P. Energy transfer: a spectroscopic ruler. *Proc. Natl Acad. Sci. USA* **58**, 719-726 (1967).

30. Lakowicz, J. R. *Principles of Fluorescence Spectroscopy*, 3rd edn (Springer, 2006).

31. Giavazzi, D., Saseendran, S., Di Maiolo, F. & Painelli, A. A comprehensive approach to exciton delocalization and energy transfer. *J. Chem. Theory Comput.* **19**, 436-447 (2023).

32. Chuang, C., Knoester, J. & Cao, J. Scaling relations and optimization of excitonic energy transfer rates between one-dimensional molecular aggregates. *J. Phys. Chem. B* **118**, 7827-7834 (2014).

33. Mathur, D. *et al.* Understanding Förster resonance energy transfer in the sheet regime with DNA brick-based dye networks. *ACS Nano* **15**, 16452-16468 (2021).

34. Halpert, J. E. *et al.* Electrostatic formation of quantum dot/J-aggregate FRET pairs in solution. *J. Phys. Chem. C* **113**, 9986-9992 (2009).

35. Chakraborty, D., Akhuli, A., Preeyanka, N. & Sarkar, M. Energy-transfer-induced enhanced valley splitting of excitonic emission of inorganic CdTe@ZnS QDs in the presence of organic

J-aggregates: a spectroscopic insight into the efficient exciton (inorganic)-exciton (organic) coupling. *J. Phys. Chem. C* **127**, 5082-5089 (2023).

36. Preeyanka, N., Dey, H., Seth, S., Rahaman, A. & Sarkar, M. Highly efficient energy transfer from a water soluble zinc silver indium sulphide quantum dot to organic J-aggregates. *Phys. Chem. Chem. Phys.* **22**, 12772-12784 (2020).
37. Qiao, Y. *et al.* Nanotubular J-aggregates and quantum dots coupled for efficient resonance excitation energy transfer. *ACS Nano* **9**, 1552-1560 (2015).
38. Zhang, Q. *et al.* Highly efficient resonant coupling of optical excitations in hybrid organic/inorganic semiconductor nanostructures. *Nat. Nanotechnol.* **2**, 555-559 (2007).
39. Mandal, S., Zhou, X., Lin, S., Yan, H. & Woodbury, N. Directed energy transfer through DNA-templated J-aggregates. *Bioconjug. Chem.* **30**, 1870-1879 (2019).
40. Banal, J. L., Kondo, T., Veneziano, R., Bathe, M. & Schlau-Cohen, G. S. Photophysics of J-aggregate-mediated energy transfer on DNA. *J. Phys. Chem. Lett.* **8**, 5827-5833 (2017).
41. Walker, B. J., Nair, G. P., Marshall, L. F., Bulović, V. & Bawendi, M. G. Narrow-band absorption-enhanced quantum dot/J-aggregate conjugates. *J. Am. Chem. Soc.* **131**, 9624-9625 (2009).
42. Freyria, F. S. *et al.* Near-infrared quantum dot emission enhanced by stabilized self-assembled J-aggregate antennas. *Nano Lett.* **17**, 7665-7674 (2017).
43. Akselrod, G. M., Walker, B. J., Tisdale, W. A., Bawendi, M. G. & Bulović, V. Twenty-fold enhancement of molecular fluorescence by coupling to a J-aggregate critically coupled resonator. *ACS Nano* **6**, 467-471 (2012).
44. Valtiner, M., Banquy, X., Kristiansen, K., Greene, G. W. & Israelachvili, J. N. The electrochemical surface forces apparatus: the effect of surface roughness, electrostatic surface potentials, and anodic oxide growth on interaction forces, and friction between dissimilar surfaces in aqueous solutions. *Langmuir* **28**, 13080-13093 (2012).

45. Pick, C., Argento, C., Drazer, G. & Frechette, J. Micropatterned charge heterogeneities via vapor deposition of aminosilanes. *Langmuir* **31**, 10725-10733 (2015).

46. Estévez-Varela, C. *et al.* Room-temperature surface exciton polaritons in colloidal J-aggregate flakes. *Adv. Opt. Mater.* **13**, e02346 (2025).

47. Rurack, K. & Spieles, M. Fluorescence quantum yields of a series of red and near-infrared dyes emitting at 600-1000 nm. *Anal. Chem.* **83**, 1232-1242 (2011).

48. Magde, D., Wong, R. & Seybold, P. G. Fluorescence quantum yields and their relation to lifetimes of rhodamine 6G and fluorescein in nine solvents: improved absolute standards for quantum yields. *Photochem. Photobiol.* **75**, 327-334 (2002).

49. Sumi, H. Theory on rates of excitation-energy transfer between molecular aggregates through distributed transition dipoles with application to the antenna system in bacterial photosynthesis. *Journal of Physical Chemistry B* **103**, 252–260 (1999).

50. Cao, J. Generalized resonance energy transfer theory: applications to vibrational energy flow in optical cavities. *J. Phys. Chem. Lett.* **13**, 10943-10951 (2022).

# Supplementary Information

## Efficient Plane-to-Point Energy Transfer from Supramolecular Excitonic Systems to Molecular Acceptors

Dimuthu H. Thanippuli Arachchi, Oliver M. Nix, Jianshu Cao, Moungi G. Bawendi*

*Department of Chemistry, Massachusetts Institute of Technology, 77 Massachusetts Avenue, Cambridge, Massachusetts, 02139, United States*

**Corresponding author: mgb@mit.edu*

**Table of Contents**

## 1. LC-MS characterization of the ATTO 590-APTES conjugate

### 1.1 Analysis of ion chromatogram peaks

The formation of the silane-functionalized ATTO 590 derivative was characterized by liquid chromatography-mass spectrometry (LC-MS). Chromatograms of the individual ATTO 590 NHS ester, APTES, and the reaction mixture following amide coupling of ATTO 590 NHS ester with APTES were first analyzed to identify their characteristic retention times and mass-to-charge ratios. The mixture exhibits new chromatographic peaks and retention shifts, consistent with amide bond formation between APTES and ATTO 590, alongside residual unreacted species.

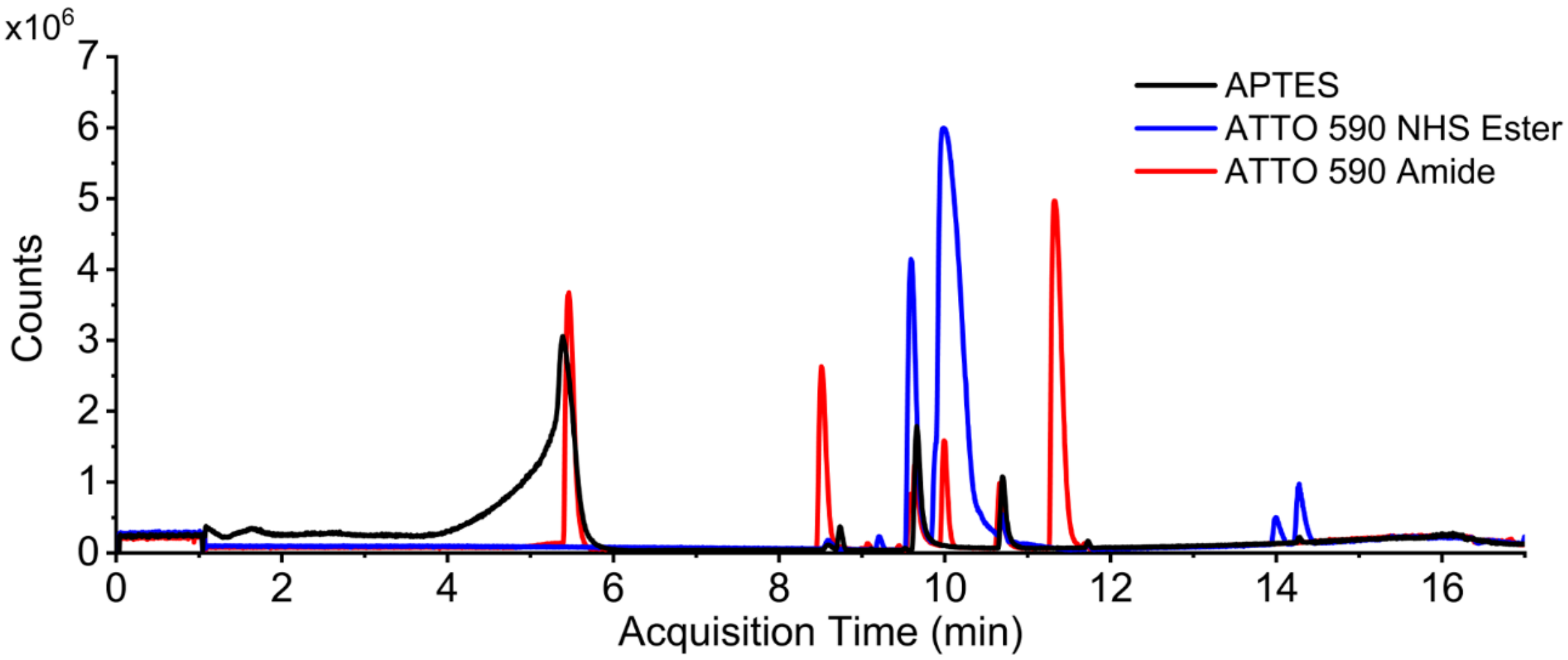

**Figure S1. LC-MS chromatographic analysis of ATTO 590-APTES conjugation.** Overlaid LC chromatograms of APTES (black), ATTO 590 NHS ester (blue), and the reaction mixture following amide coupling of ATTO 590 NHS ester with APTES (ATTO 590 amide, red).

### 1.2 Analysis of ion chromatogram of APTES

LC-MS analysis of APTES was performed to identify the characteristic signals of the silane precursor and distinguish them from products formed following reaction with ATTO 590 NHS ester. The APTES chromatogram shows a major peak at 5.39 min (peak 1), assigned to APTES, together with a second component at 8.74 min (peak 2), assigned to an APTES-derived dimer, and two additional minor peaks attributed to organic impurities (peaks 3 and 4).

The mass spectrum of peak 1 contains the protonated molecular ion of APTES at m/z 222.1518, consistent with the calculated $[M+H]^+$ value of 222.1520, as well as a fragment at m/z 176.1095 attributed to in-source fragmentation involving loss of an ethoxy group as ethanol. Peak 2 exhibits a molecular ion at m/z 426.2713, consistent with the assigned APTES-derived dimer $[M+H]^+$ = 426.2702. The remaining chromatographic features at 9.66 and 10.69 min exhibit ions centered at m/z 230.2491 and 258.28, respectively, and were assigned as potential organic impurities rather than APTES-derived species used in the subsequent conjugation analysis.

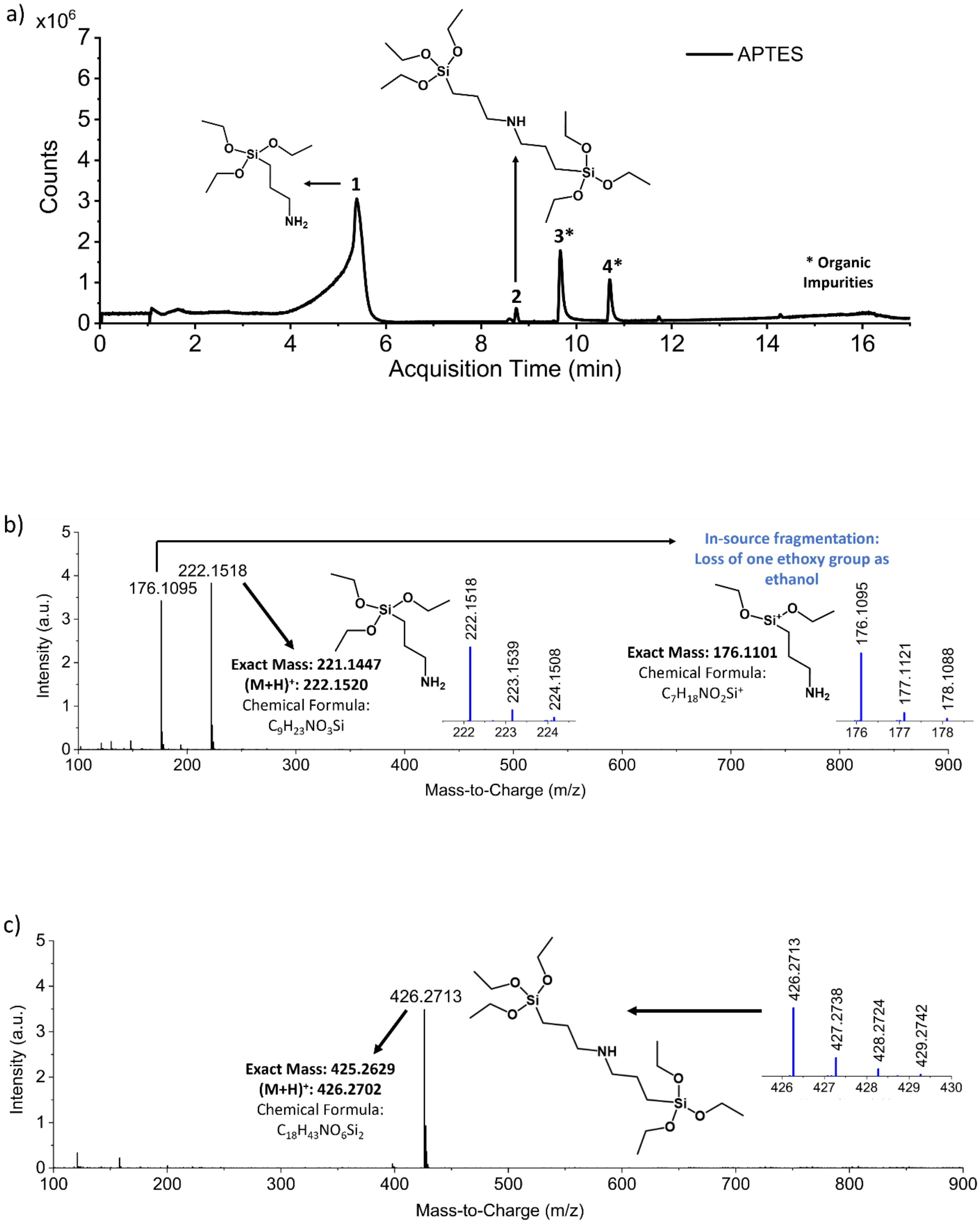

a)
x10^6
Counts
Acquisition Time (min)
APTES
1
2
3*
4*
* Organic Impurities
NH2
NH
Si
b)
Intensity (a.u.)
Mass-to-Charge (m/z)
176.1095
222.1518
Exact Mass: 221.1447
(M+H)+: 222.1520
Chemical Formula: C9H23NO3Si
In-source fragmentation: Loss of one ethoxy group as ethanol
Exact Mass: 176.1101
Chemical Formula: C7H18NO2Si+
223.1539
224.1508
177.1121
178.1088
c)
426.2713
Exact Mass: 425.2629
(M+H)+: 426.2702
Chemical Formula: C18H43NO6Si2
427.2738
428.2724
429.2742

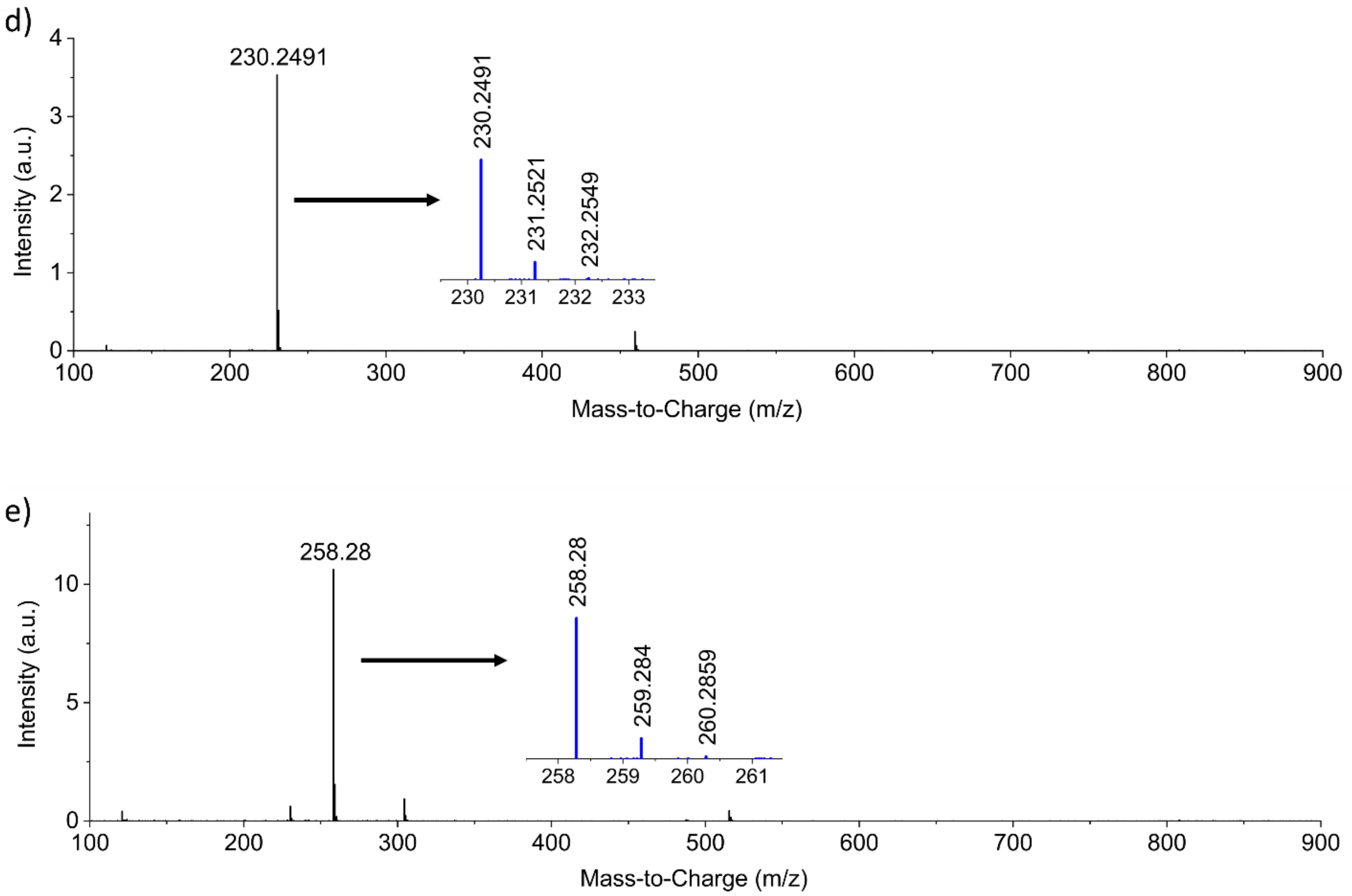

**Figure S2. LC-MS characterization of APTES.** a) Chromatogram of APTES. b) APTES, 5.39 min ($[M+H]^+$ at m/z 222.1518; in-source fragment at m/z 176.1095). c) APTES-derived dimer, 8.74 min ($[M+H]^+$ at m/z 426.2713). d) Unidentified impurity, 9.66 min (m/z 230.2491). e) Unidentified impurity, 10.69 min (m/z 258.28).

## 1.3 Analysis of ion chromatogram of ATTO 590 NHS Ester

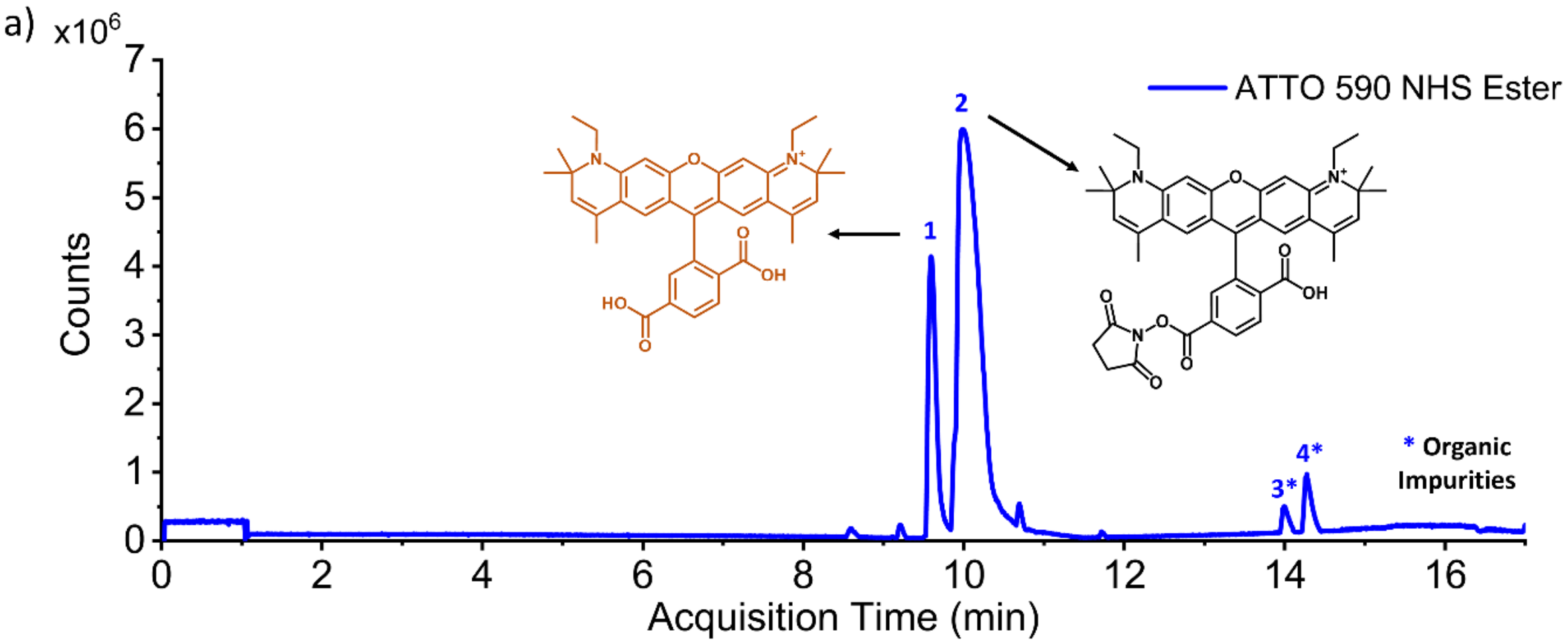

b)

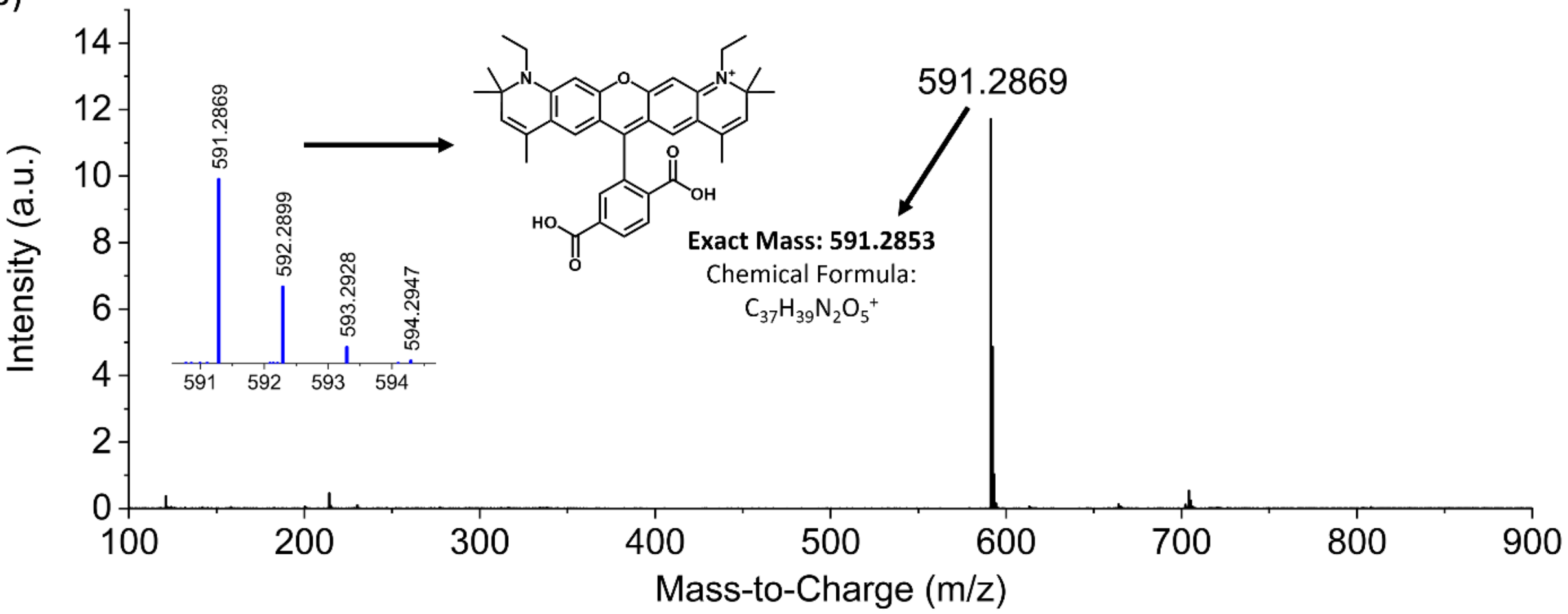
591.2869
591.2869
592.2899
593.2928
594.2947
591
592
593
594
Exact Mass: 591.2853
Chemical Formula:
$C_{37}H_{39}N_2O_5^+$
Intensity (a.u.)
Mass-to-Charge (m/z)

c)

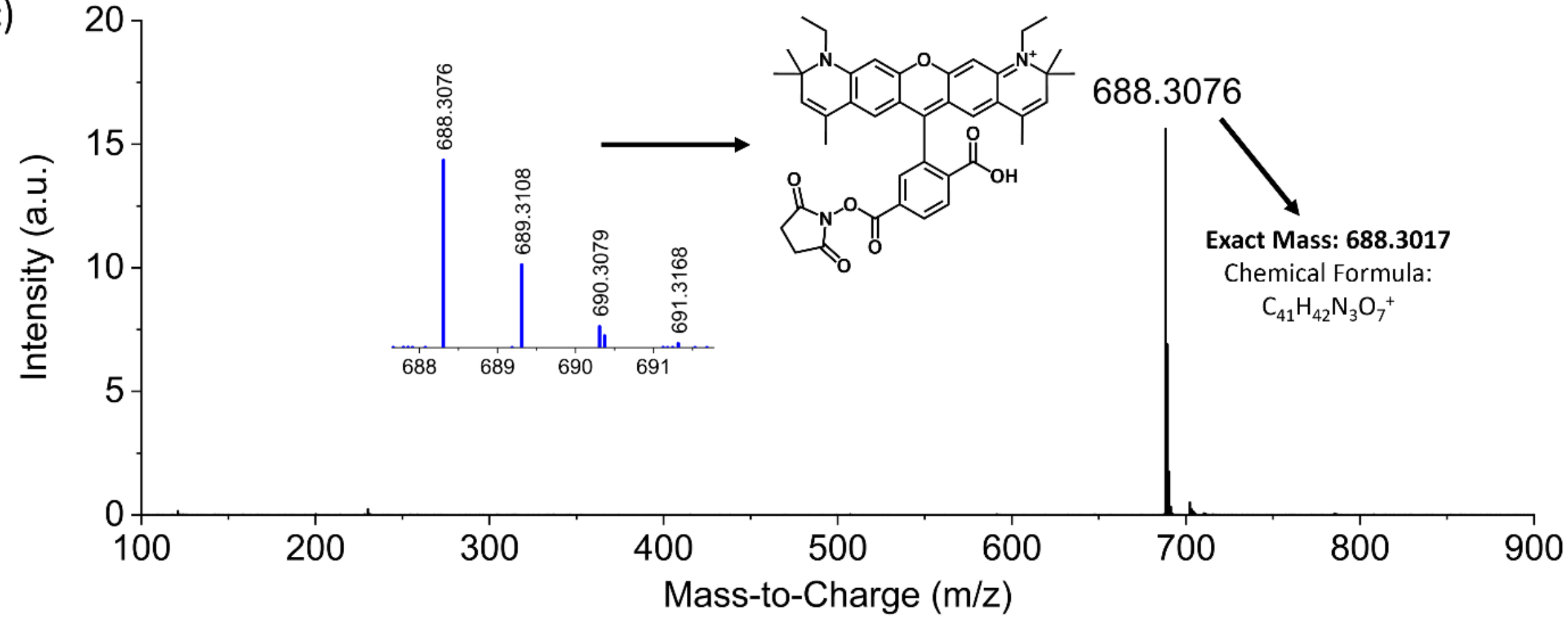
688.3076
688.3076
689.3108
690.3079
691.3168
688
689
690
691
Exact Mass: 688.3017
Chemical Formula:
$C_{41}H_{42}N_3O_7^+$
Intensity (a.u.)
Mass-to-Charge (m/z)

d)

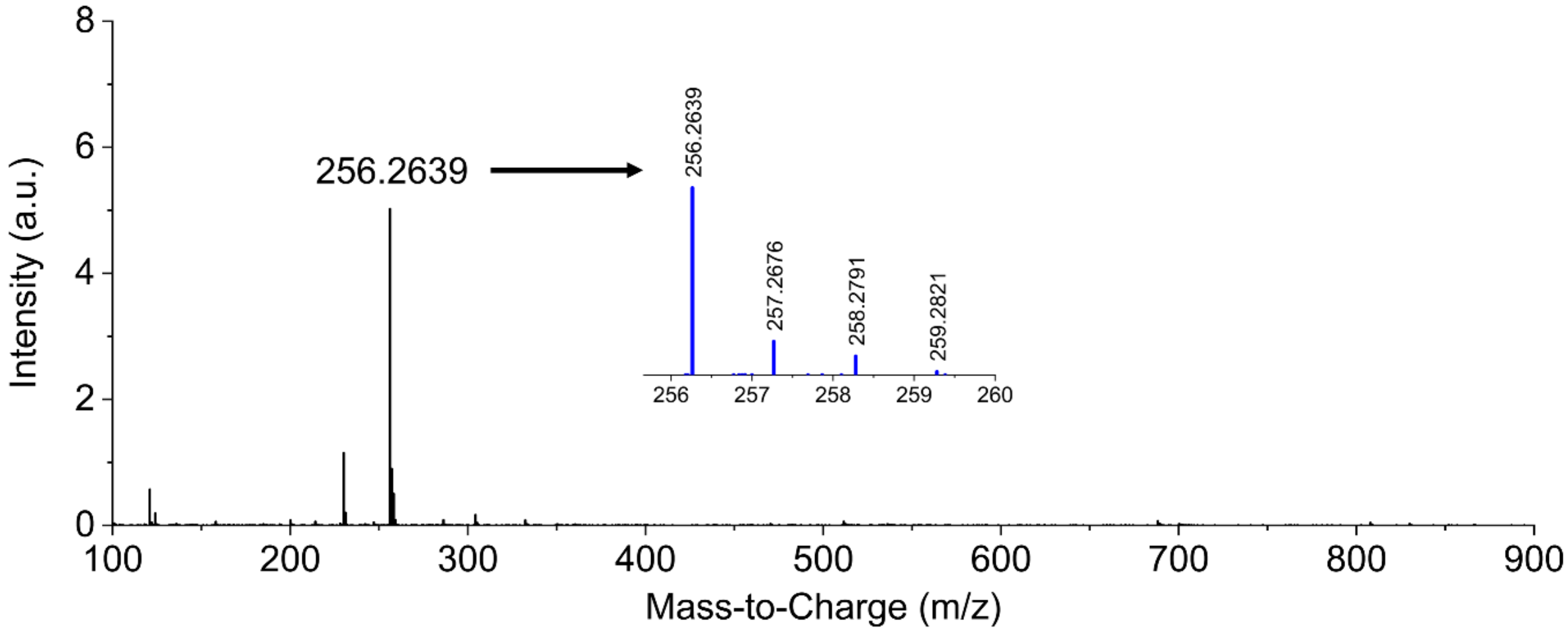
256.2639
256.2639
257.2676
258.2791
259.2821
256
257
258
259
260
Intensity (a.u.)
Mass-to-Charge (m/z)

e)

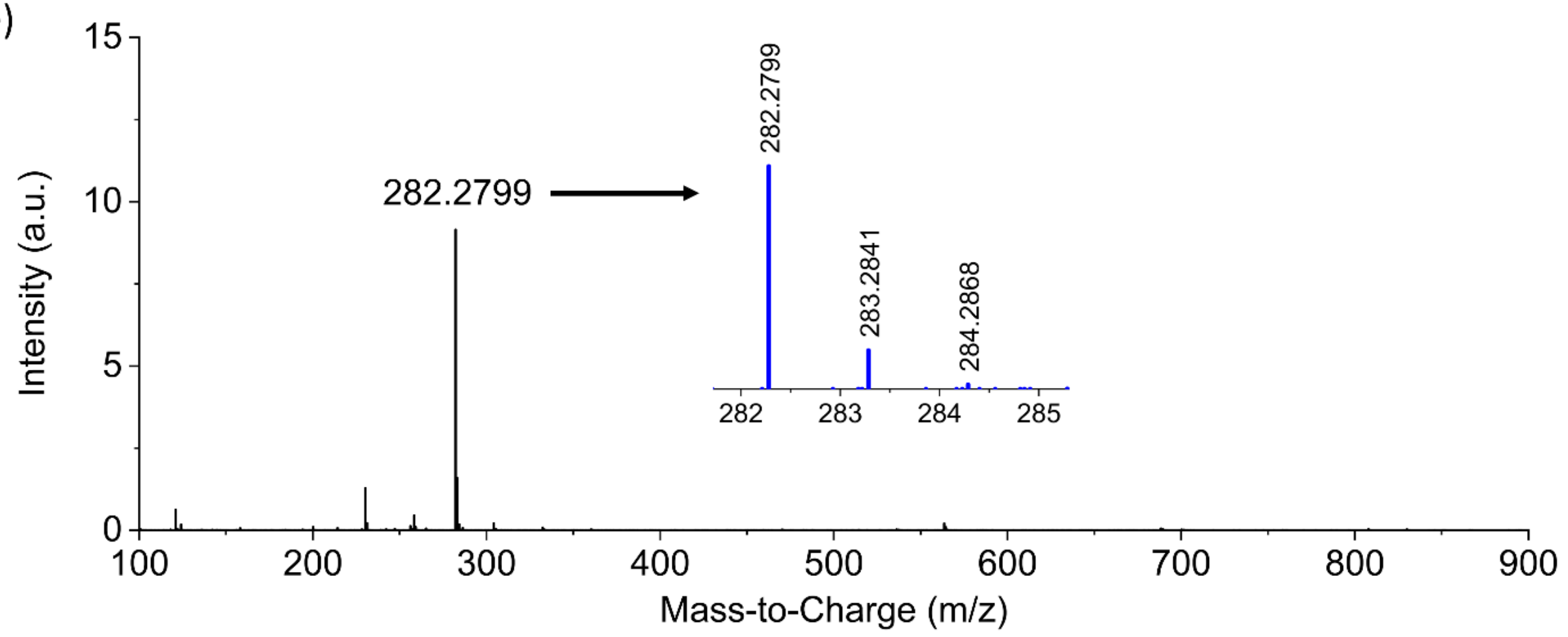

**Figure S3. LC-MS characterization of ATTO 590 NHS ester.** a) Chromatogram of ATTO 590 NHS ester. b) ATTO 590 acid, 9.59 min (m/z 591.2869). c) ATTO 590 NHS ester, 9.98 min (m/z 688.3076). d) Unidentified impurity, 13.99 min (m/z 256.2639). e) Unidentified impurity, 14.28 min (m/z 282.2799).

LC-MS analysis of ATTO 590 NHS ester identified two major dye-derived species (Fig. S3). The peak at 9.59 min was assigned to hydrolyzed ATTO 590 acid, exhibiting an ion at m/z 591.2869, consistent with the calculated mass of 591.2853. The major peak at 9.98 min was assigned to intact ATTO 590 NHS ester, with an ion at m/z 688.3076 (calculated mass 688.3017). Two additional minor peaks at 13.99 and 14.28 min, with major ions at m/z 256.2639 and 282.2799, respectively, could not be assigned to ATTO 590 NHS ester or its expected derivatives and were therefore attributed to unidentified organic impurities.

### 1.4 Analysis of ion chromatogram of ATTO 590 Amide (APTES + NHS Ester Mixture)

LC-MS analysis of the reaction mixture following coupling of ATTO 590 NHS ester with APTES revealed a series of ATTO 590-APTES species corresponding to different degrees of silane hydrolysis (Fig. S4). The peak at 5.46 min was assigned to excess APTES. Peaks at 8.51, 9.06, 9.99, and 11.32 min were assigned to ATTO 590-amide-propylsilanetriol (m/z 710.3268), ATTO 590-amide-propylethoxysilanediol (m/z 738.3576), ATTO 590-amide-propyldiethoxysilanol (m/z 766.3896), and ATTO 590-amide-propyltriethoxysilane (m/z 794.4207), respectively. ATTO 590 acid was observed at 9.59 min (m/z 591.2854). Peaks at 9.64 and 10.66 min were assigned to impurities originating from the APTES starting material. The observation of ATTO 590-APTES species across multiple silane hydrolysis states confirms formation of the silane-functionalized ATTO 590 amide conjugate.

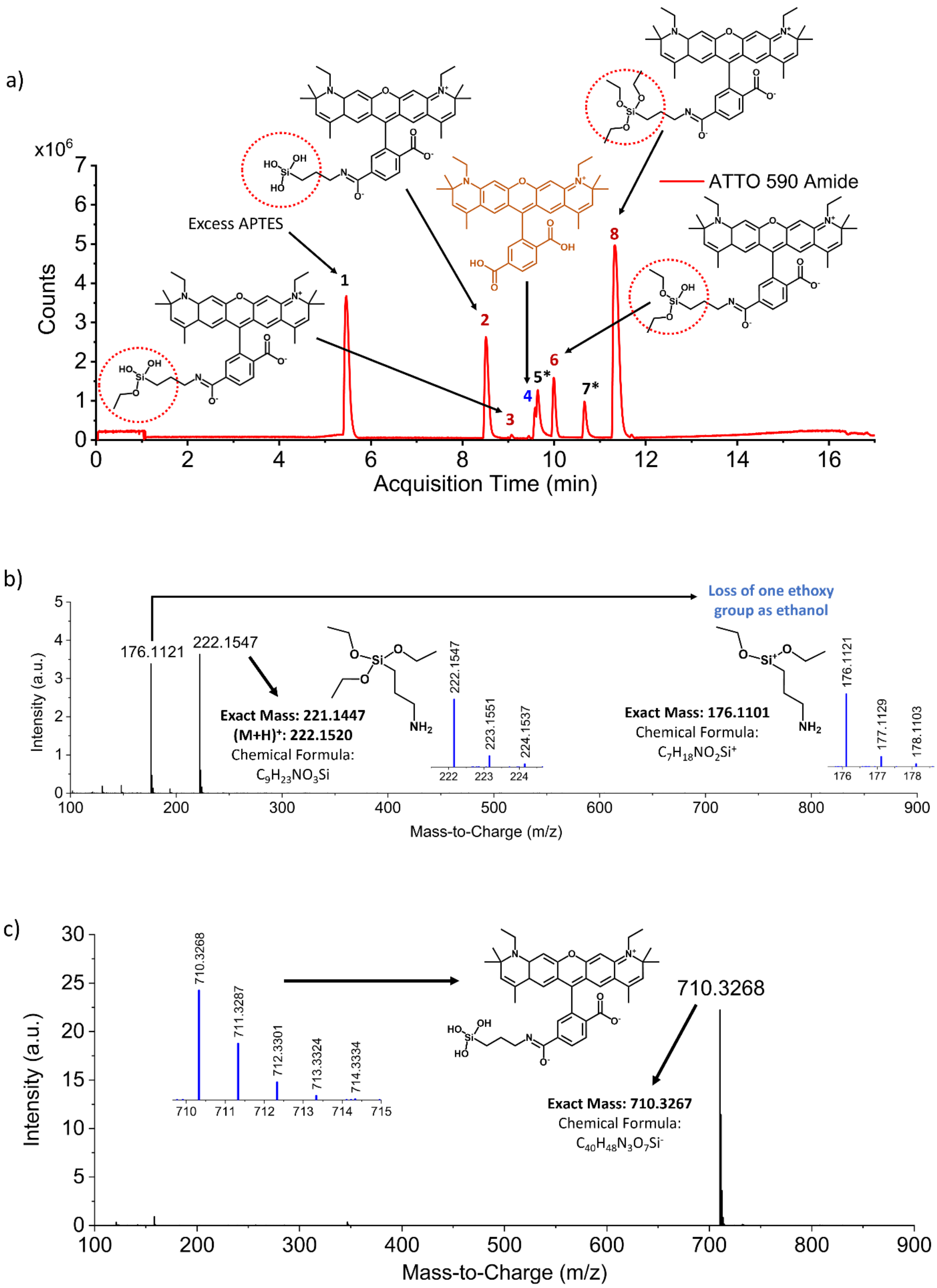
a)
x10^6
Counts
Acquisition Time (min)
ATTO 590 Amide
Excess APTES
1
2
3
4
5*
6
7*
8
b)
Intensity (a.u.)
Mass-to-Charge (m/z)
176.1121
222.1547
Loss of one ethoxy group as ethanol
Exact Mass: 221.1447
(M+H)+: 222.1520
Chemical Formula: C9H23NO3Si
Exact Mass: 176.1101
Chemical Formula: C7H18NO2Si+
222.1547
223.1551
224.1537
176.1121
177.1129
178.1103
c)
Intensity (a.u.)
Mass-to-Charge (m/z)
710.3268
711.3287
712.3301
713.3324
714.3334
710.3268
Exact Mass: 710.3267
Chemical Formula: C40H48N3O7Si-

d)

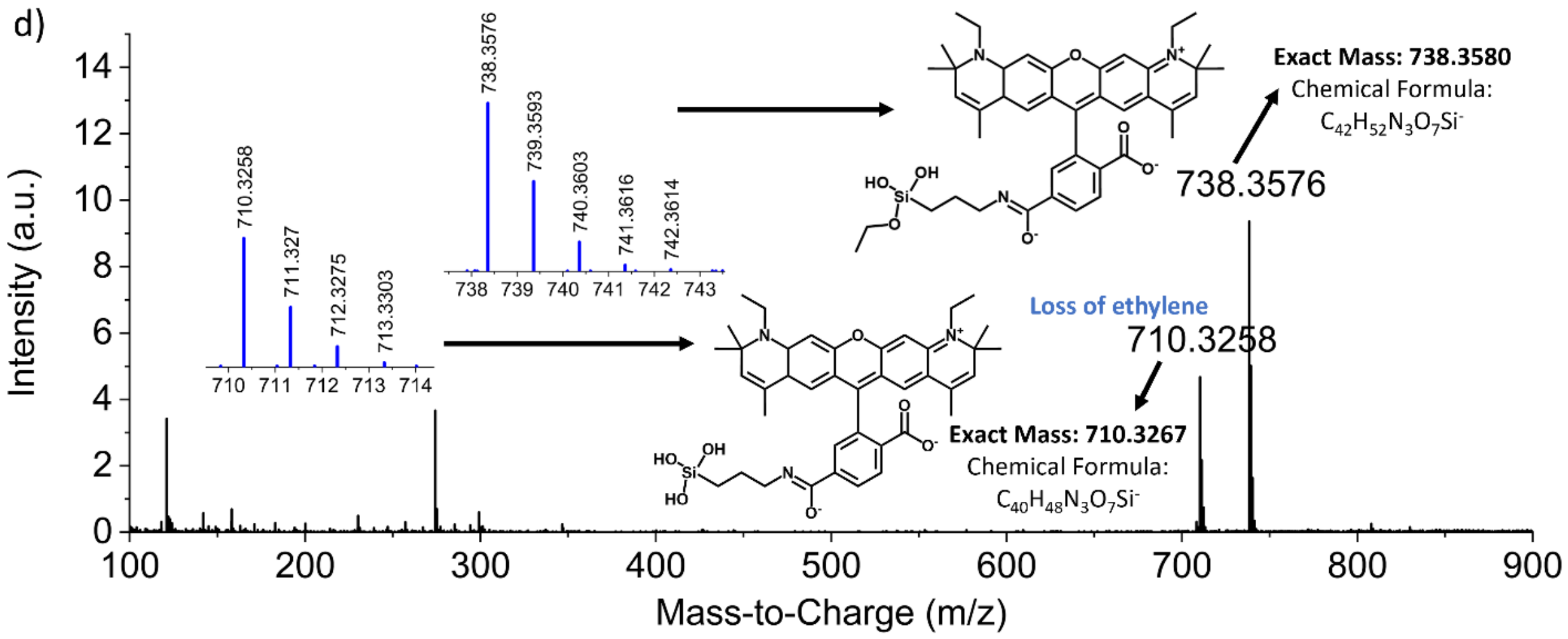

Intensity (a.u.)
Mass-to-Charge (m/z)
710.3258
711.327
712.3275
713.3303
738.3576
739.3593
740.3603
741.3616
742.3614
Exact Mass: 738.3580
Chemical Formula:
$C_{42}H_{52}N_3O_7Si^-$
738.3576
Loss of ethylene
710.3258
Exact Mass: 710.3267
Chemical Formula:
$C_{40}H_{48}N_3O_7Si^-$

e)

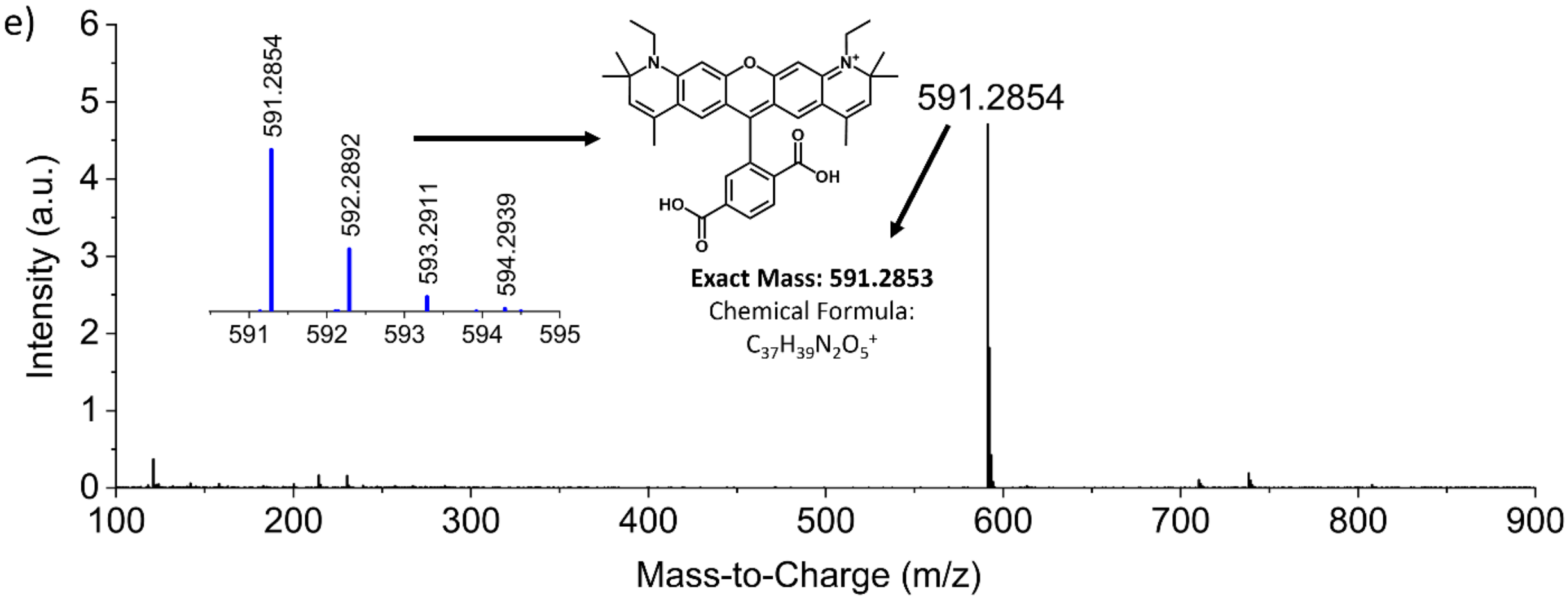

Intensity (a.u.)
Mass-to-Charge (m/z)
591.2854
592.2892
593.2911
594.2939
591.2854
Exact Mass: 591.2853
Chemical Formula:
$C_{37}H_{39}N_2O_5^+$

f)

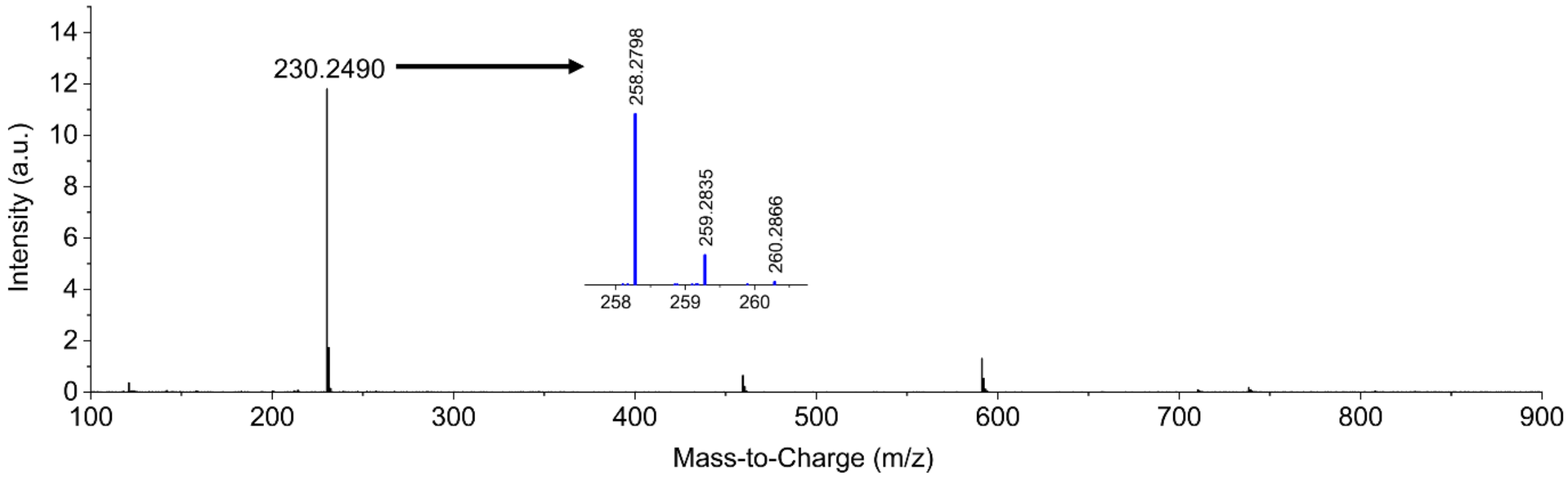

Intensity (a.u.)
Mass-to-Charge (m/z)
230.2490
258.2798
259.2835
260.2866

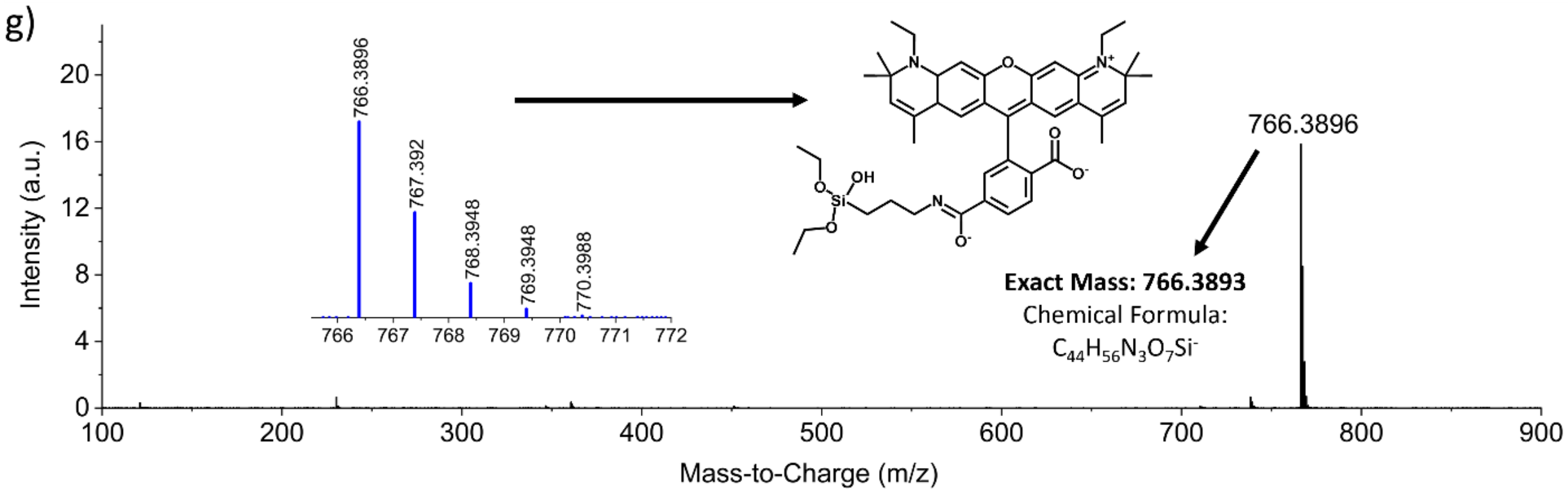

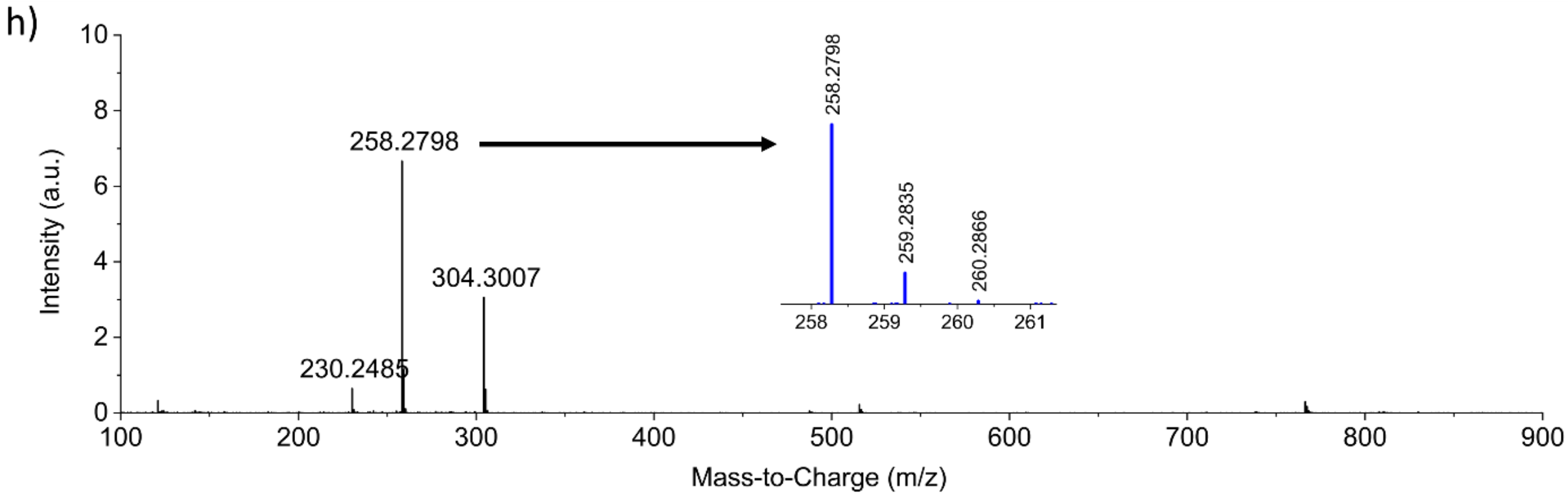

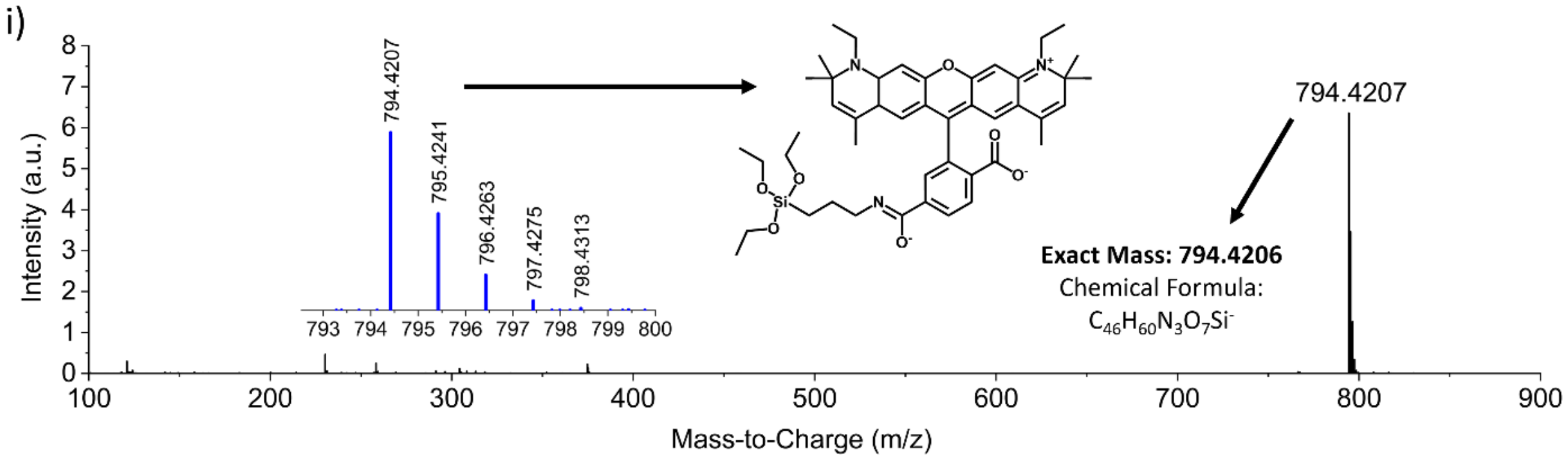

**Figure S4. LC-MS characterization of the ATTO 590-APTES conjugation reaction**. a) Chromatogram of the reaction mixture. b) Excess APTES, 5.46 min. c) ATTO 590-amide-propylsilanetriol, 8.51 min (m/z 710.3268). d) ATTO 590 amide-propylethoxysilanediol, 9.06 min (m/z 738.3576). e) ATTO 590 acid, 9.59 min (m/z 591.2854). f) APTES associated impurity, 9.64 min. g) ATTO 590-amide-propyldiethoxysilanol, 9.99 min (m/z 766.3896). h) APTES-associated impurity, 10.66 min. i) ATTO 590-amide-propyltriethoxysilane, 11.32 min (m/z 794.4207).

LC-MS peak intensities and areas were used only to establish the presence of the corresponding species and were not interpreted as measures of relative concentration, as differences in ionization efficiency, ion suppression, fragmentation, and matrix effects can produce substantially different detector responses among species.

## 2. Determination of TDBC/ATTO 590 molar ratios

To determine the total TDBC concentration, TDBC J-aggregates were first separated from unbound monomers by centrifugation, and the sedimented aggregates were dissolved in methanol to recover the constituent TDBC monomers prior to standard-addition analysis. The native TDBC concentration was then obtained from the x-intercept of the linear absorbance versus added TDBC concentration plot. Similarly, each silica-encapsulated TDBC J-aggregate-ATTO 590 construct was centrifuged and disassembled into its constituent chromophores prior to determining the ATTO 590 content by standard addition with successive additions of known ATTO 590 concentrations. The independently determined TDBC and ATTO 590 concentrations were then used to calculate TDBC/ATTO 590 molar ratios of approximately 90:1, 63:1, 44:1, and 19:1.

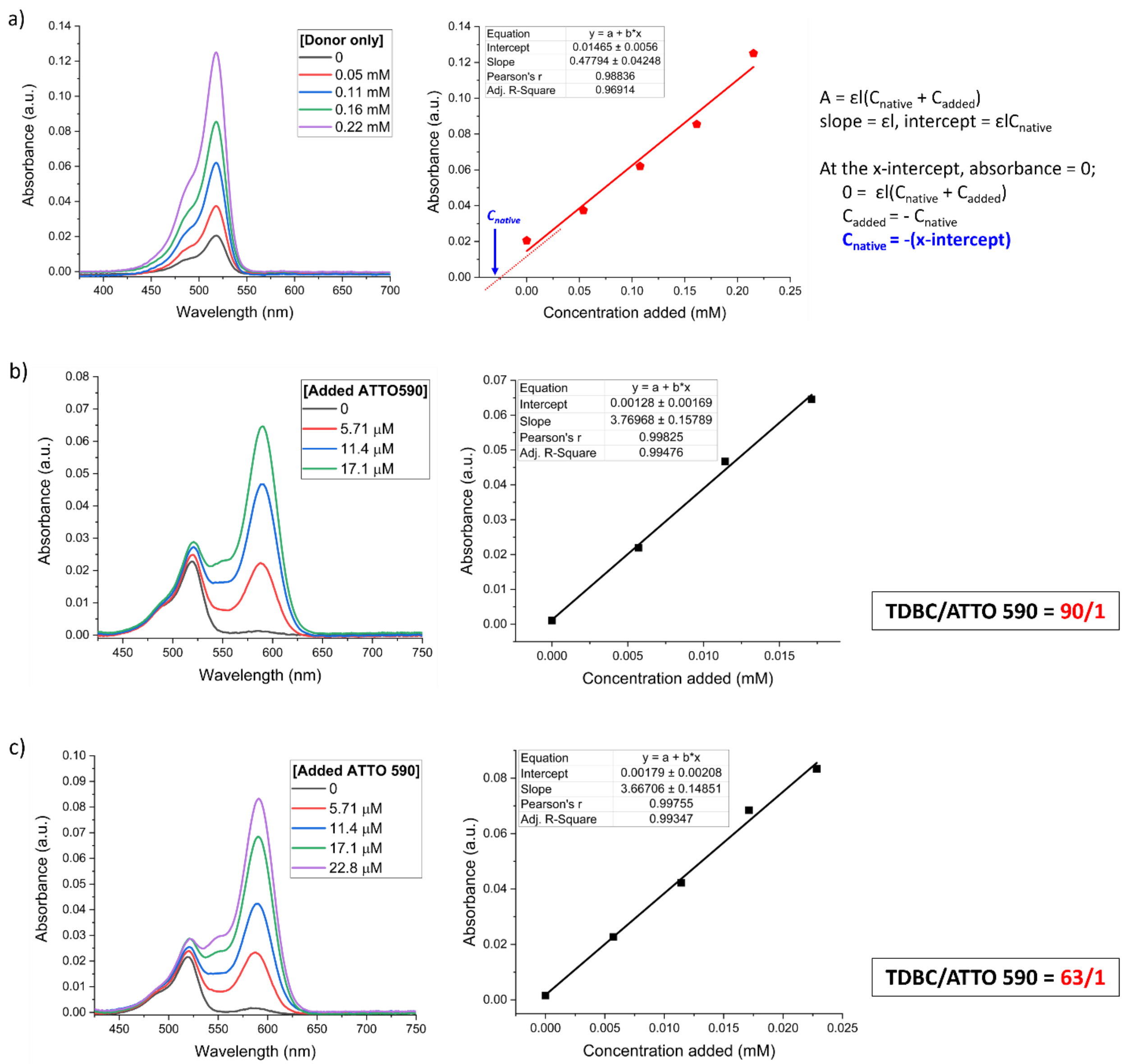

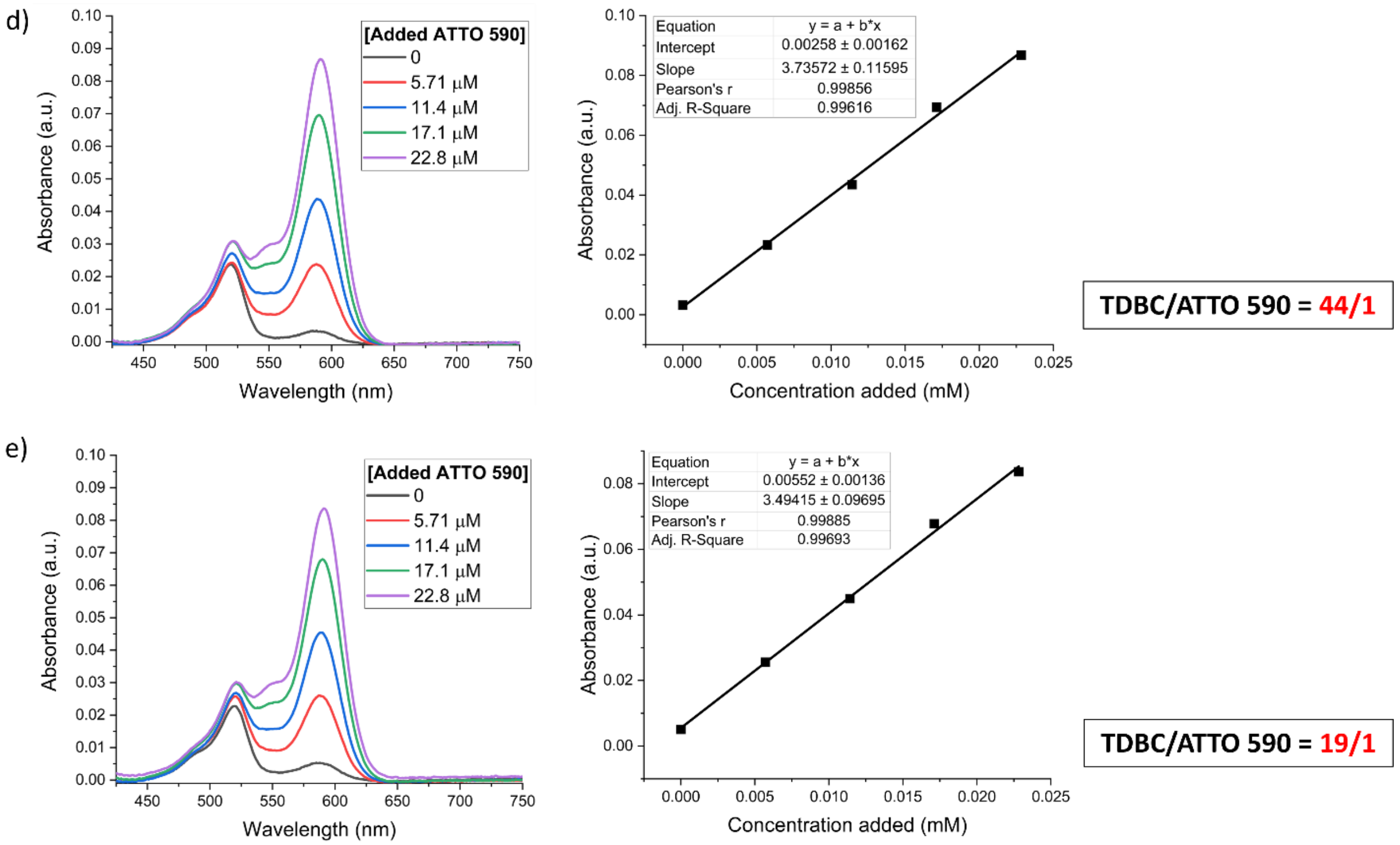

**Figure S5. Determination of TDBC/ATTO 590 molar ratios by standard addition.** a) Absorption spectra of the disassembled TDBC J-aggregate donor following successive additions of known TDBC concentrations (left) and the corresponding standard addition plot (right), from which the native TDBC monomer concentration was obtained from the x-intercept. b-e) Absorption spectra of the disassembled donor-acceptor constructs following successive additions of known ATTO 590 concentrations (left) and corresponding standard-addition plots (right), yielding TDBC/ATTO 590 molar ratios of 90:1 (b), 63:1 (c), 44:1 (d), and 19:1 (e).

## 3. Spectral overlap between TDBC J-aggregates and ATTO 590

The normalized emission spectrum of the TDBC J-aggregate donor overlaps substantially with the absorption spectrum of ATTO 590-APTES, providing the spectral requirement for Förster-type energy transfer. The calculated spectral overlap integral was $J = 3.57 \times 10^{15}\ M^{-1}\ cm^{-1}\ nm^4$.

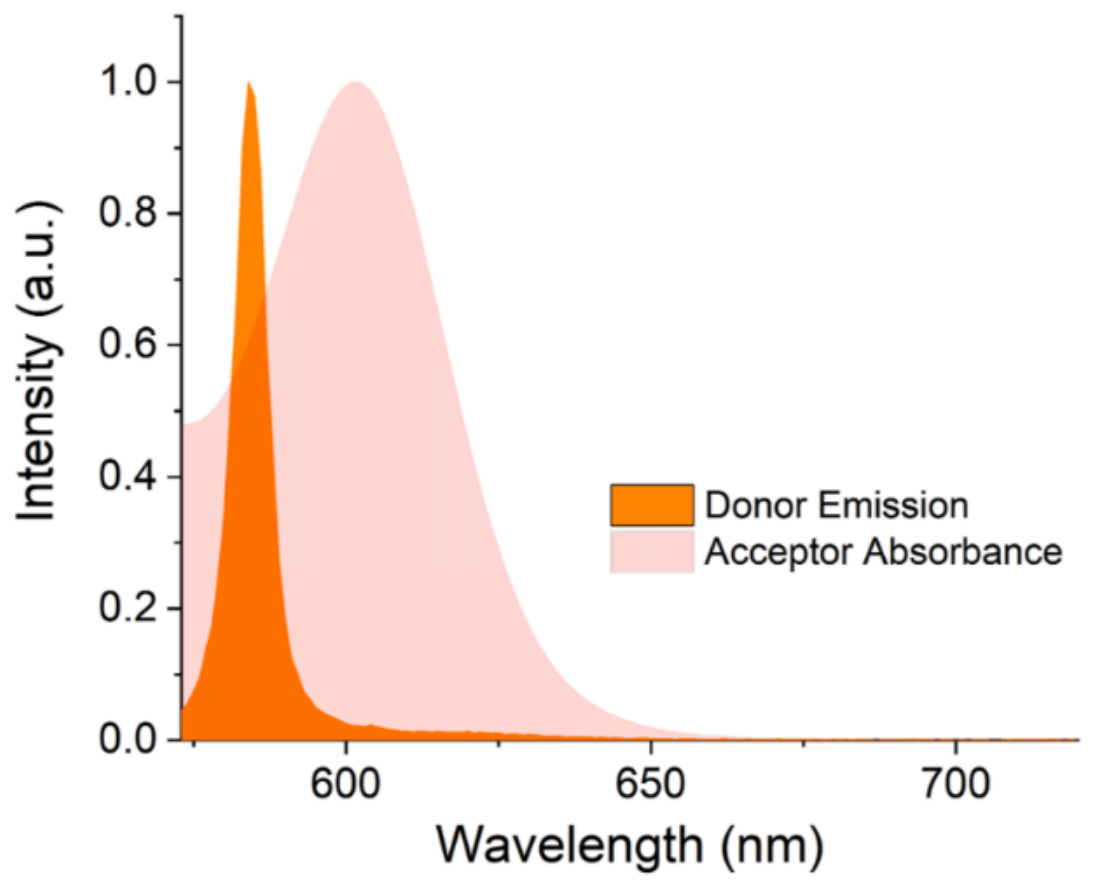

**Figure S6. Spectral overlap between TDBC J-aggregates and ATTO 590.** Normalized TDBC J-aggregate donor emission and ATTO 590 acceptor absorption spectra showing the spectral overlap relevant to donor-to-acceptor energy transfer.

## 4. Photoluminescence quantum yield measurements of silica-encapsulated TDBC J-aggregates

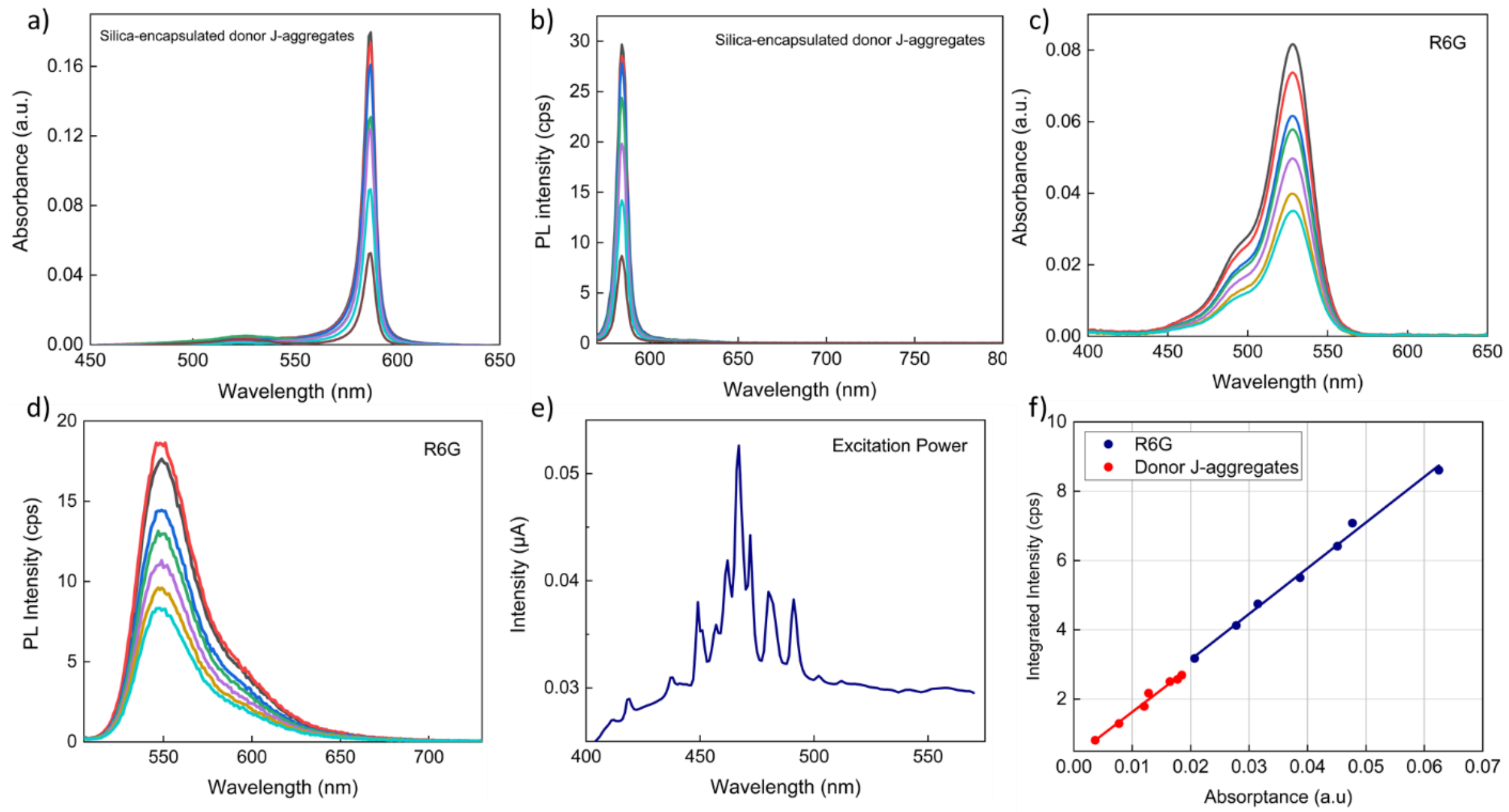

**Figure S7. Calculation of photoluminescence quantum yield of silica-encapsulated TDBC Donor J-Aggregates.** a) Absorbance and b) emission spectra of TDBC donor J-aggregates at different concentrations. c) Absorbance and d) emission spectra of R6G solutions in methanol at different concentrations. e) Excitation power of the fluorimeter as a function of wavelength. f) Plot of integrated emission as a function of absorptance for R6G (blue) and TDBC donor J-aggregates (red).

$$\text{QY (silica-encapsulated TDBC donor J-aggregates)} = \frac{1.29\cdot 10^{8}}{1.32\cdot 10^{8}} \cdot \left(\frac{1.333}{1.329}\right)^{2} \cdot \frac{0.0307}{0.0297} \cdot 93\% \ = 94\%\,,$$

where the ratios represent the slopes of the calibration plots, refractive indices of water and methanol, and powers of excitation at 500 nm for rhodamine 6G and at 565 nm for the silica-encapsulated TDBC J-aggregates, respectively.

## 5. Photoluminescence quantum yield measurements of ATTO 590-APTES conjugate

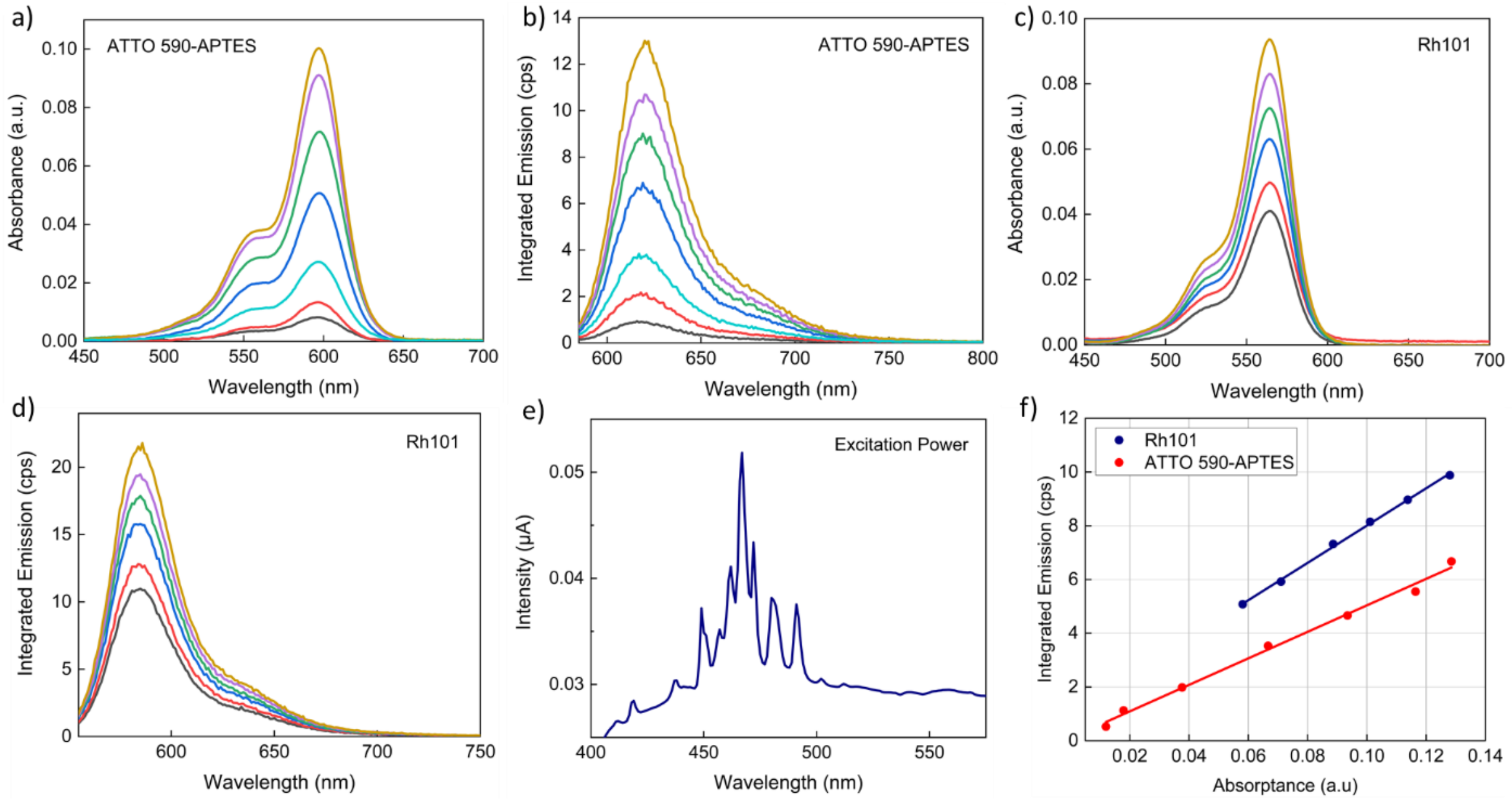

**Figure S8. Calculation of photoluminescence quantum yield of ATTO 590-APTES conjugate. a)** Absorbance and **b)** emission spectra of ATTO 590-APTES at different concentrations. **c)** Absorbance and **d)** emission spectra of Rh101 solutions in ethanol at different concentrations. **e)** Excitation power of the fluorimeter as a function of wavelength. **f)** Plot of integrated emission as a function of absorptance for Rh101 (blue) and ATTO 590-APTES (red).

$$\text{QY (ATTO 590-APTES conjugate)} = \frac{4.93 \cdot 10^{7}}{6.94 \cdot 10^{7}} \cdot \left(\frac{1.406}{1.361}\right)^{2} \cdot \frac{0.0292}{0.0290} \cdot 91\% \ = 70\% \,,$$

where the ratios represent the slopes of the calibration plots, refractive indices of water/DMSO mixture and ethanol, and powers of excitation at 550 nm for rhodamine 101 and at 580 nm for the ATTO 590-APTES conjugate, respectively.

## 6. Energy transfer efficiency calculations

### 6.1 Energy transfer efficiency based on donor quenching

Energy-transfer (ET) efficiencies were independently estimated from donor quenching, acceptor sensitization, and donor lifetime shortening.[1] For the steady-state donor-quenching analysis, the remaining TDBC donor contribution in the donor-acceptor construct was obtained from spectral decomposition to remove the overlapping ATTO 590 emission (Fig. 4a). The intensity-based ET efficiency ($E_{ET,int}$) was calculated at the donor emission maximum ($\lambda_{max}$=586 nm) according to,

$$E_{ET,int} = 1 - \frac{I_{DA}}{I_D}$$

where $I_D$ and $I_{DA}$ are the donor fluorescence intensities in the absence and presence of acceptor, respectively, yielding an ET efficiency of approximately **79%**. An area-based ET efficiency ($E_{ET,area}$) was independently obtained from the integrated donor emission,

$$E_{ET,area} = 1 - \frac{\int I_{DA}(\lambda)d\lambda}{\int I_D(\lambda)d\lambda}$$

giving an ET efficiency of approximately **75%**.

### 6.2 Energy transfer efficiency based on acceptor sensitization

Acceptor sensitization was quantified from the spectrally decomposed donor-acceptor emission. The sensitized ATTO 590 contribution, $A_{Sens}$, was obtained after subtraction of the directly excited acceptor contribution, while $A_{Donor\ remaining}$ represents the residual donor emission. The relative sensitization efficiency ($E_{Sens}$) was calculated as,

$$E_{Sens} = \frac{A_{Sens}}{A_{Sens} + A_{Donor\ remaining}}$$

giving approximately **75%**. Because the donor and acceptor have different photoluminescence quantum yields, the emission areas were additionally corrected by their respective quantum yields,

$$E_{Sens,corr} = \frac{A_{Sens}/\Phi_A}{A_{Sens}/\Phi_A + A_{Donor\ remaining}/\Phi_D}$$

where $\Phi_A$ and $\Phi_D$ are the photoluminescence quantum yields of ATTO 590-APTES conjugate and the silica-encapsulated TDBC J-aggregate donor, respectively. This correction yielded an acceptor-sensitization-derived ET efficiency ($E_{Sens,\ corr}$) of approximately **81%**.

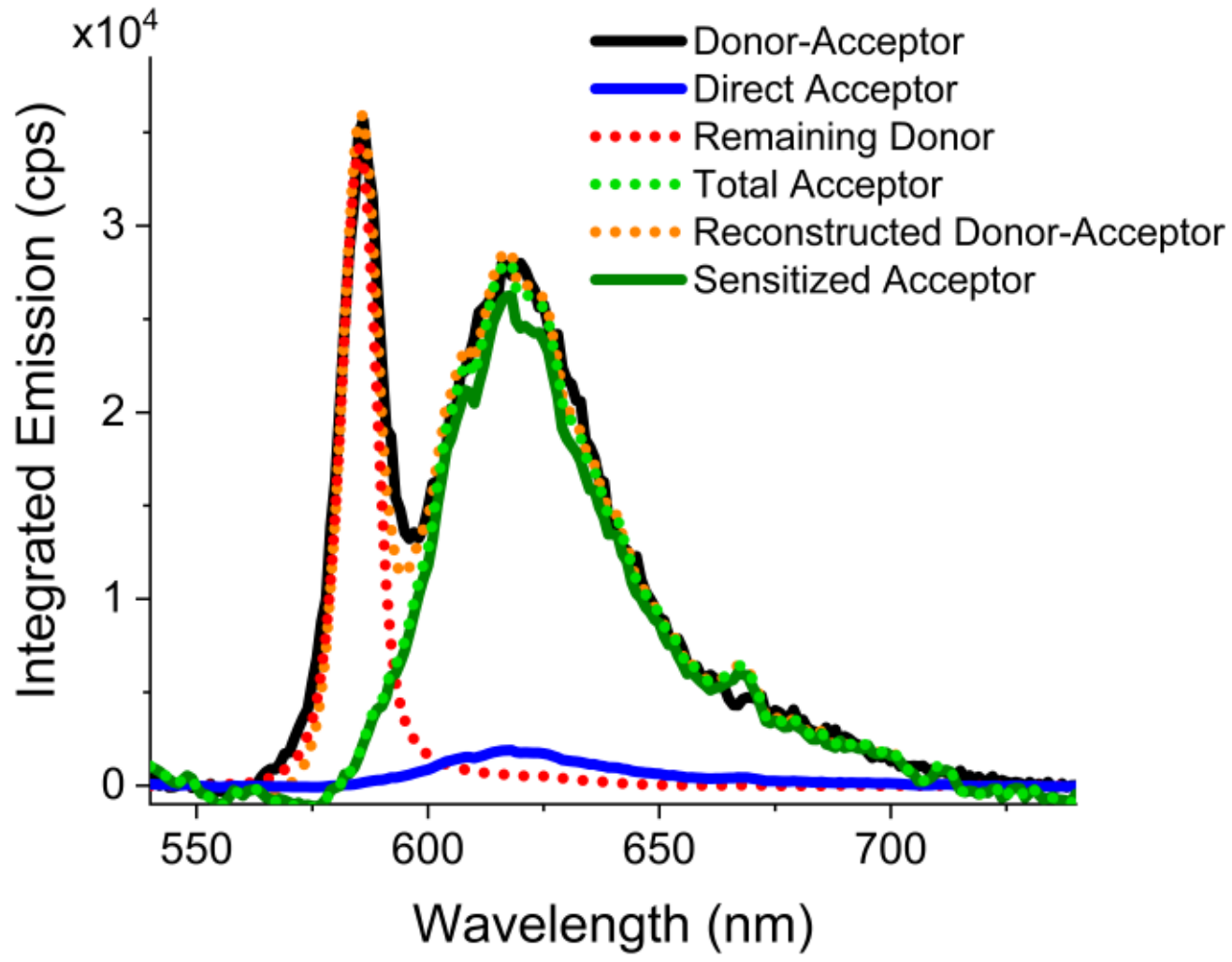

**Figure S9. Spectral decomposition of the donor-acceptor construct used to quantify sensitized ATTO 590 emission.** The steady-state photoluminescence spectrum of the donor-acceptor construct (black) was decomposed into the remaining donor emission (red, dotted) and the total acceptor contribution (green, dotted), with the reconstructed donor-acceptor spectrum shown for comparison (orange, dotted). The directly excited ATTO 590 emission is shown separately (blue), and the difference between the total and direct acceptor contributions corresponds to the sensitized ATTO 590 emission arising from energy transfer (green, solid). Integrated donor and sensitized-acceptor emission areas were calculated over 550-650 nm and 580-720 nm, respectively, and corrected for their respective photoluminescence quantum yields for estimation of the energy-transfer efficiency.

### 6.3 Energy transfer efficiency based on donor lifetimes

Time-resolved photoluminescence measurements showed a pronounced shortening of the TDBC J-aggregate donor lifetime following coupling to ATTO 590. The donor lifetime decreased from 206.9 ± 1.2 ps for the donor-only sample to 93.3 ± 2.2 ps for the donor-acceptor construct. The ET efficiency was determined from the shortening of the donor excited-state lifetime using,

$$E_{ET,\tau} = 1 - \frac{\tau_{DA}}{\tau_D}$$

where $\tau_D$ and $\tau_{DA}$ are the donor lifetimes in the absence and presence of acceptor, respectively, yielding an ET efficiency of approximately **55%**.

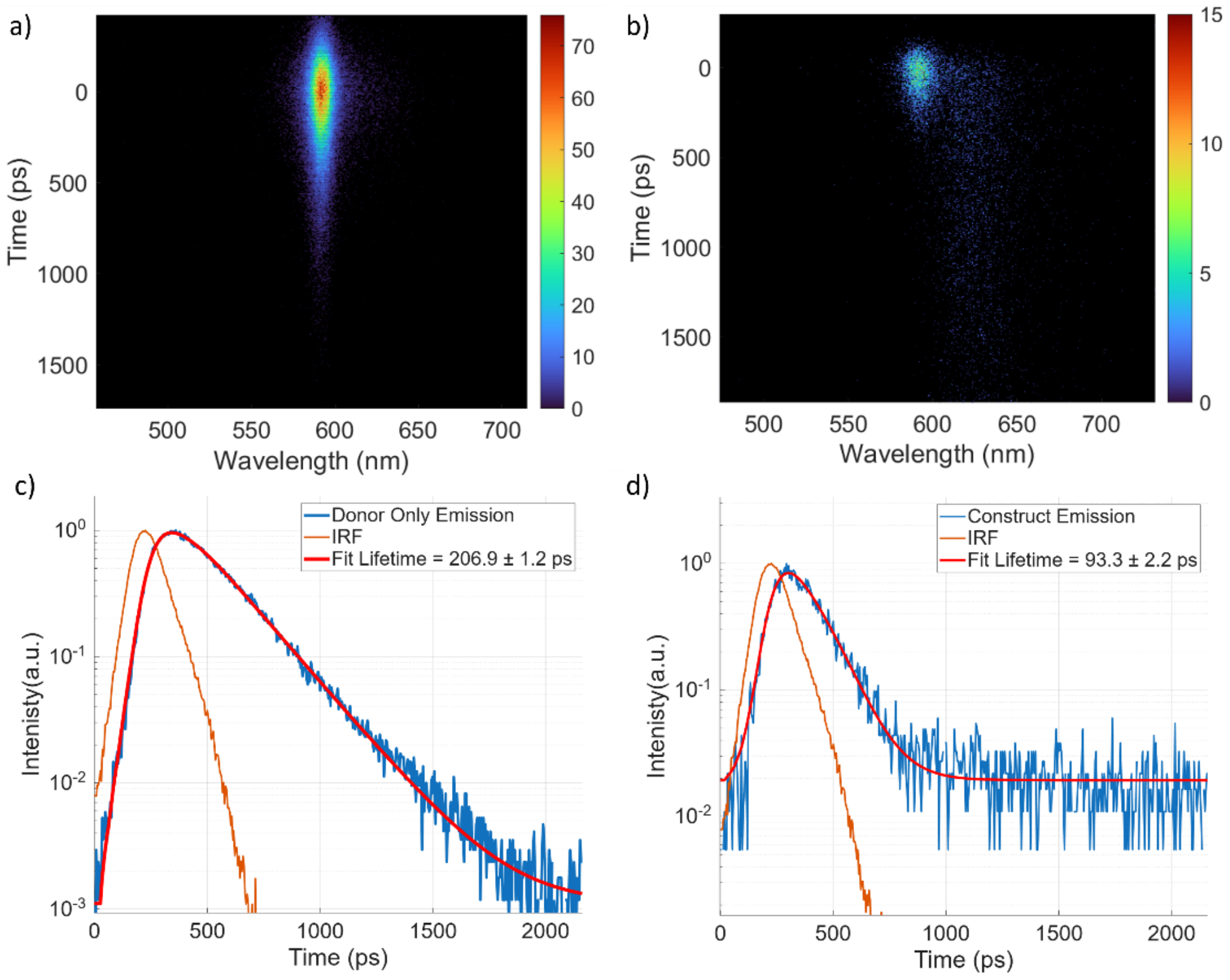

**Figure S10. Time-resolved photoluminescence analysis of donor-only and donor-acceptor samples.** a),b) Time-resolved emission maps of the silica-encapsulated TDBC donor only sample and the Silica-encapsulated TDBC J-aggregate-ATTO 590 donor-acceptor construct, respectively. c),d) Corresponding donor photoluminescence decay traces and fits, yielding lifetimes of 206.9 ps for the donor-only sample and 93.3 ± 2.2 ps for the donor-acceptor construct.

## 7. Derivation of the plane-to-point Förster Resonance Energy Transfer (FRET) distance dependence

### 7.1 Effective dipole coupling

Consider dipole coupling between donor (D) and acceptor (A)

$$T_{DA} = \frac{(\vec{\mu}_A \vec{\mu}_D) - 3(\hat{R}\vec{\mu}_D)(\hat{R}\vec{\mu}_A)}{R^3} \qquad (1)$$

where $\vec{R}$ is the vector connecting the donor and acceptor, $\hat{R} = \vec{R}/R$ is the corresponding unit vector, and $\vec{\mu}_D$ and $\vec{\mu}_A$ are the donor and acceptor dipoles, respectively.

In the experimental system, the acceptor dye molecules are independent, with no excitonic coupling or coherence between different acceptors, whereas the monomers within the J-aggregate sheet are coherently coupled over a length scale smaller than the lateral dimensions of the physical aggregate. In addition, the monomer transition dipoles are ordered such that $\vec{\mu}_D$ is a constant over the entire 2D sheet. Thus, evaluation of the FRET rate involves the following product of dipole couplings,

$$|T_{DA}|^2 = \frac{1}{R^3R'^3}[9(\hat{R}\vec{\mu}_D)(\hat{R}\vec{\mu}_A)(\hat{R}'\vec{\mu}_D)(\hat{R}'\vec{\mu}_A) + (\vec{\mu}_A\vec{\mu}_D) - 3(\hat{R}\vec{\mu}_A)(\hat{R}\vec{\mu}_D)(\vec{\mu}_A\vec{\mu}_D) - 3(\hat{R}'\vec{\mu}_A)(\hat{R}'\vec{\mu}_D)(\vec{\mu}_A\vec{\mu}_D)] \quad (2)$$

where $\vec{R}$ and $\overrightarrow{R'}$ are associated with two distinct molecules on the sheet. Since the acceptor is isotropic and randomly oriented, we can average the transition dipole moment of the acceptor molecule, $\vec{\mu}_A$, over the solid angle uniformly and obtain

$$\langle \vec{\mu}_A \cdot \vec{\mu}_A \rangle_\Omega = (\frac{|\mu_A|^2}{3}) I$$

where the subscript Ω indicates the orientational average and I is an identity matrix. Then, the orientational averaging yields

$$\langle |T_{DA}|^2 \rangle_\Omega = \frac{\mu_A^2}{R^3R'^3} [3(\hat{R}\hat{R}')(\hat{R}\hat{\mu}_D)(\hat{R}'\vec{\mu}_D) + \frac{\mu_D^2}{3} - (\hat{R}\vec{\mu}_D)(\hat{R}\vec{\mu}_D) - (\hat{R}'\overrightarrow{\mu_D})(\hat{R}'\vec{\mu}_D)] \quad (3)$$

which represents the effective dipole-coupling strength between an isotropic acceptor and an ordered and coherent pair of monomers in the J-aggregate plane (Fig. S11).

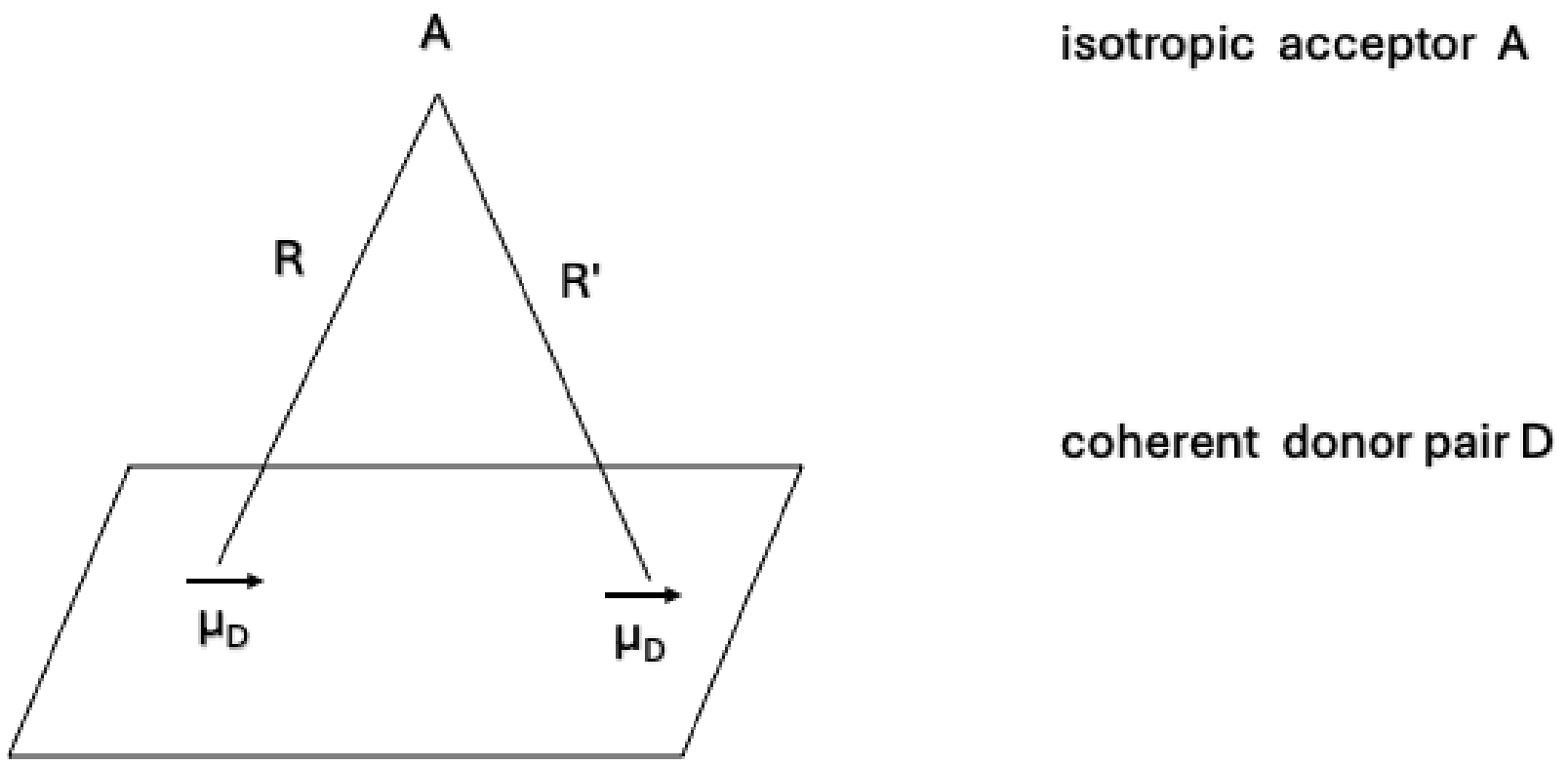

**Figure S11.** Effective dipole coupling between an isotropically-oriented acceptor and a pair of coherently-coupled donors with aligned dipoles.

### 7.2 Infinite spatial coherence in J-aggregate

Now we consider the infinite coherent limit of the J-aggregate, where the exciton wavefunction covers the entire sheet. In this limit, we adopt Eq. (3) with R ≠ R′, such that the dipole-coupling intensity, $J_{cpl}^2$, becomes

$$J_{cpl}^2 = \mu_A^2 \rho_A \int dS dS' \frac{1}{R^3 R'^3} [3(\hat{R}\hat{R}')(\hat{R}'\vec{\mu}_D)(\hat{R}\vec{\mu}_D) - (\hat{R}\vec{\mu}_D)^2 - (\hat{R}'\vec{\mu}_D)^2 + \frac{\mu_D^2}{3}] \quad (4)$$

where $\rho_A$ is the areal number density of acceptors. Here, the donors are a 2D J-aggregate sheets, with an area much larger than the donor-acceptor separation (z). The three terms in the above integral are evaluated as follows:

$$\int dS \frac{1}{R^3} = \int dr \frac{2\pi r}{(r^2 + z^2)^{3/2}} = \frac{2\pi}{z}$$

$$\int dS \frac{x^2}{R^5} = \int dr \int d\theta \frac{\cos^2\theta\, r^3}{(r^2 + z^2)^{5/2}} = \frac{\pi}{2} \int \frac{r^2}{(r^2 + z^2)^{3/2}} dr^2 = \frac{2\pi}{3z}$$

$$\int dS \int dS' (xx' + yy' + z^2)(xx')(\frac{1}{R^5 R'^5}) = [\int dS \frac{x^2}{R^5}]^2 = [\frac{2\pi}{3z}]^2$$

Putting all the terms into Eq. (4), we find the overall dipole-coupling intensity as

$$J_{cpl}^2 = \mu_A^2 \mu_D^2 \rho_A [3(\tfrac{2\pi}{3z})^2 - 2(\tfrac{2\pi}{3z})(\tfrac{2\pi}{z}) + \tfrac{1}{3}(\tfrac{2\pi}{z})^2] = 0 \quad (5)$$

which vanishes in the limit of infinite coherent length.

The surprising prediction of vanishing dipole coupling is consistent with the 1D wire calculation published earlier.[2] To understand this, we consider the simple case of a dipole interacting with a 1D J-aggregate of length $n_{coh}$. At the far field, the 1D aggregate acts as a collective dipole of $\vec{\mu}_{coh} = n_{coh}\vec{\mu}$. At the near-field, the head-to-tail arrangement in the J-aggregate chain leads to cancellation of charges, leaving residual charge at the two ends. The amplitude of the residual charge scales as $1/\sqrt{n_{coh}}$ so the effective coupling with the point dipole reduces monotonically with coherence length $n_{coh}$. Therefore, in the infinite coherent-length limit, the effective coupling vanishes.[2]

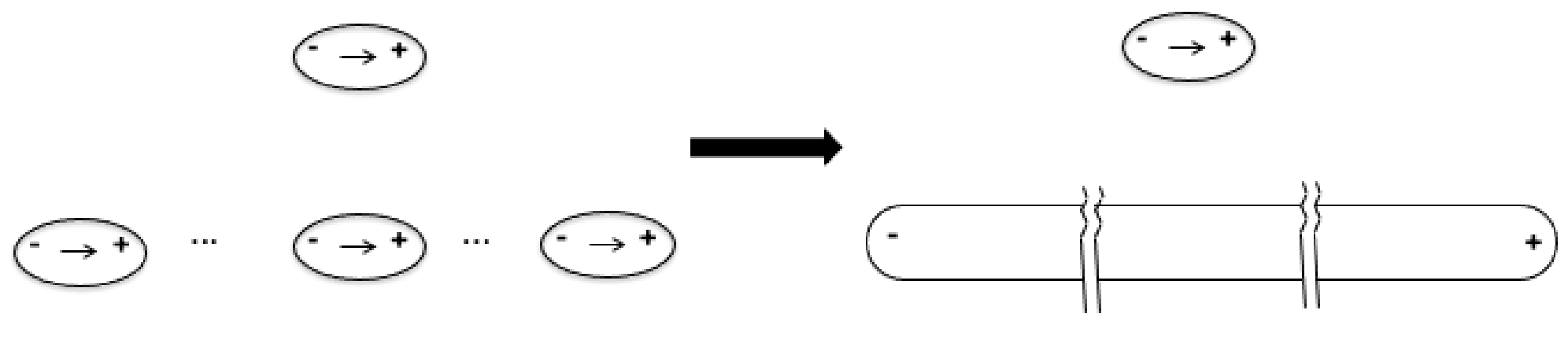

**Figure S12. Effective dipole coupling between a dye molecule and a linear J-aggregate.** At the near-field, the head-to-tail arrangement in the 1D J-aggregate leads to cancellation of charges, leaving residual charge at the two ends. The amplitude of the residual charge scales as $1/\sqrt{n_{coh}}$ so the effective dipole coupling reduces monotonically with coherence length $n_{coh}$.

### 7.3 Distance scaling for finite spatial coherence

As shown above, if the in-plane dipoles are arranged in J-configuration, the collective dipole will vanish for infinite coherent size. Yet, for finite coherent size, the dipoles in either J or H aggregates add up collectively. Specifically, since the size of the aggregate sheet is much larger than the exciton coherent size $n_{coh}$, we can treat the coherently coupled molecules as a cluster. Each cluster is coupled to the acceptor molecule collectively with the effective dipole of a cluster, $\vec{\mu}_{coh} = \vec{\mu} n_{coh}$, but there is no quantum coherence between different clusters.[3] Effectively, we can approximate coherent energy transfer as an incoherent energy-transfer process with a scaled dipole for the aggregate sheet, such that we set $\overrightarrow{R} = \vec{R}'$ in Eq. (3).

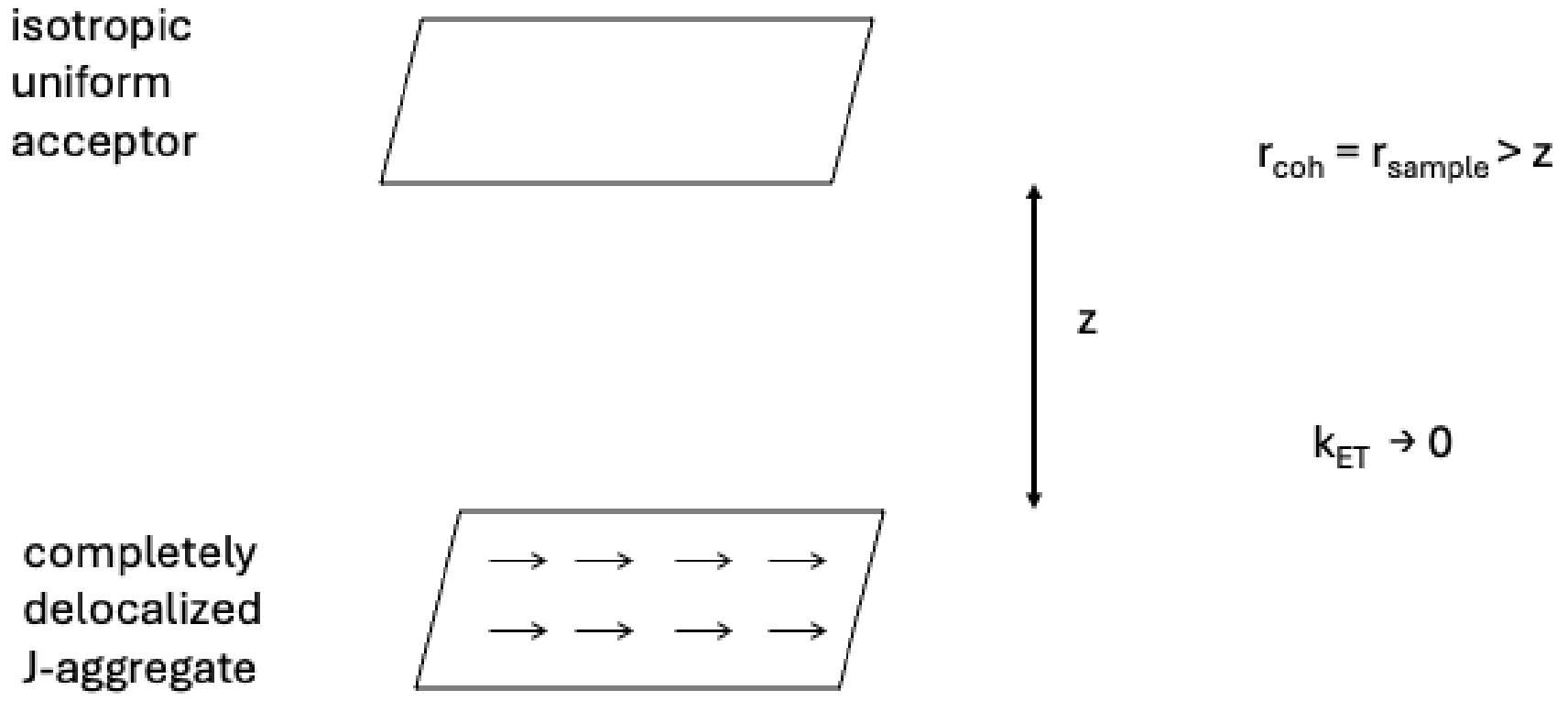

**Figure S13. Energy transfer between isotropically-oriented acceptors randomly distributed on a plane and a coherent 2D J-aggregate.** The coherent length of the J-aggregate covers the entire plane, which is much larger than the inter-plane separation z. Ignoring the edge effect, the transfer rate vanishes because of destructive interference between aligned dipoles in the J-aggregate.

We further assume that the acceptors are randomly and uniformly distributed on a plane a distance z away from the aggregate sheet, with number density $\rho_A$. The area of the z-plane is the same as the area of the sheet and is much larger than the inter-plane separation z or the coherent size $n_{coh}$, so we can ignore the edge effect and assume an infinite 2D plane. Under these conditions, the effective dipole-coupling strength can be written as

area integral

$$J_{cpl}^2 = \rho_A \int \langle |T_{DA}|^2 \rangle_\Omega \, dS \quad (6)$$

where dS is the surface-area integral.

$$\int \langle |T_{DA}|^2 \rangle_\Omega \, dS = \int dS \, \frac{\mu_A^2}{R^6} \left[ (\hat{R}\vec{\mu}_{coh})^2 + \frac{\mu_{coh}^2}{3} \right]$$

For simplicity, we convert the Cartesian coordinate to the polar coordinate, giving

cartesian polar coordinates

$$\vec{R} = (x, y, z) = (r\cos\theta, r\sin\theta, z)$$

We then evaluate the two integrals, giving

$$\int dS \frac{1}{R^6} = \int \frac{2\pi r dr}{(r^2+z^2)^3} = \frac{\pi}{2}\frac{1}{z^4}$$

and

$$\int dS \frac{\vec{x}^2}{R^8} = \int dr d\theta \frac{\cos^2\theta\, r^3}{(r^2+z^2)^4} = \frac{\pi}{2}\int \frac{dr^2}{(r^2+z^2)^4} = \frac{\pi}{12}\frac{1}{z^4}$$

Finally, putting all the pieces together, we have

$$J_{cpl}^2 = \rho_A \mu_A^2\, \mu_{coh}^2 \left[\frac{\pi}{12}\frac{1}{z^4} + \frac{1}{3}\frac{\pi}{2}\frac{1}{z^4}\right]$$

$$= \rho_A \mu_A^2\, \mu_{coh}^2\, \frac{\pi}{4}\frac{1}{z^4} \quad (7)$$

The above result predicts the inverse fourth-power scaling with the inter-plane separation z for the inter-plane energy transfer. This prediction is essentially the classical energy transfer rate law between two parallel planes. The quantum effect is accounted for by the renormalization of the in-plane dipole $\mu_{coh}$.[3]

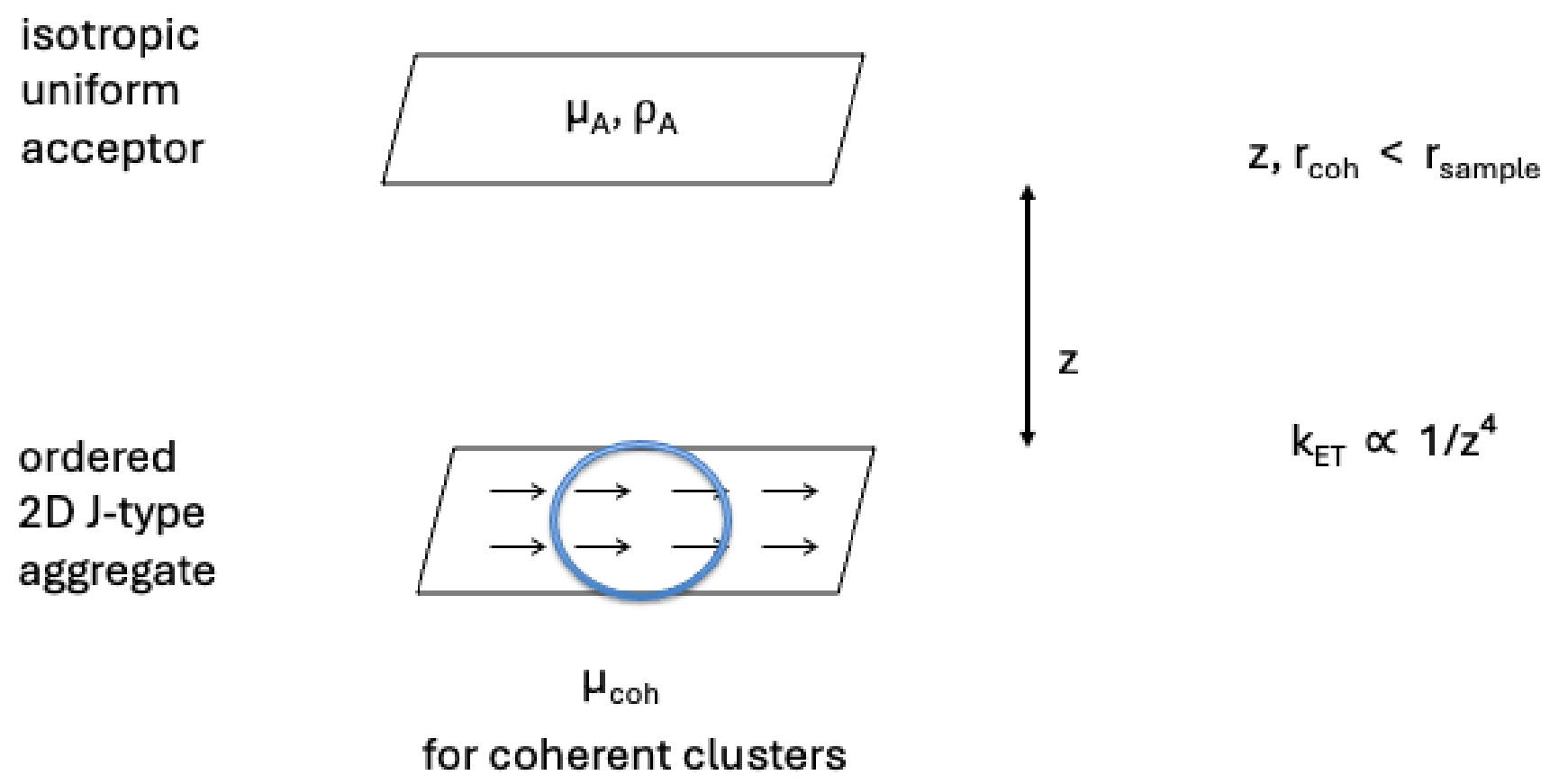

**Figure S14. Energy transfer between isotropically-oriented acceptors randomly distributed on a plane and a 2D J-aggregate of finite coherent size.** Both the coherent length and inter-plane separation are much smaller than the size of the plane. The finite coherent size $n_{coh}$ leads to renormalization of the collective transition dipole, $\vec{\mu}_{coh} = \vec{\mu} n_{coh}$. Then, the energy transfer reduces to incoherent inter-plane transfer with the renormalized dipole and inverse $z^4$ scaling.

## 7.4 Energy transfer rate for finite spatial coherence

Eq (7) contains the dipolar and geometrical contribution to the coupling but not yet the energetic resonance between donor emission and acceptor absorption. The dipole-dipole interaction energy ($V_{DA}$), contains the electrostatic prefactor. For one donor-acceptor pair, the actual interaction energy is,

$$V_{DA} = \frac{1}{4\pi\varepsilon_0\varepsilon_r} T_{DA}$$

where $\varepsilon_0$ is vacuum permittivity, $\varepsilon_r$ is the relative optical permittivity of the surrounding medium. At optical frequencies, the dielectric response is commonly approximated $\varepsilon_r \approx n^2$ where n is the effective refractive index and therefore, summing over the whole acceptor plane;

$$\langle |V_{DA}|^2 \rangle_\Omega = \frac{1}{(4\pi\varepsilon_0 n^2)^2} \langle |T_{DA}|^2 \rangle_\Omega$$

$$\rho_A \int \langle |V_{DA}|^2 \rangle_\Omega \, dS = V_{cpl}^2 = \frac{1}{(4\pi\varepsilon_0 n^2)^2} \rho_A \int \langle |T_{DA}|^2 \rangle_\Omega \, dS$$

$$V_{cpl}^2 = \frac{1}{(4\pi\varepsilon_0 n^2)^2} J_{cpl}^2 \quad (8)$$

Using Eq (7);

$$V_{cpl}^2 = \frac{1}{(4\pi\varepsilon_0 n^2)^2} \rho_A \mu_A^2 \, \mu_{coh}^2 \, \frac{\pi}{4z^4} \quad (9)$$

$V_{cpl}^2$ represents the sum of the squared pair wise interaction energies over the acceptor plane.

Eq (9) describes the magnitude and geometry of the electronic coupling but does not yet account for energetic resonance between donor emission and acceptor absorption. Let $\mathrm{g}_D(\mathrm{E})$ and $\mathrm{g}_A(\mathrm{E})$ denote normalized donor emission and acceptor absorption line shape functions.

$$\int \mathrm{g}_D(\mathrm{E}) \, dE = 1 \qquad \int \mathrm{g}_A(\mathrm{E}) \, dE = 1$$

The corresponding normalized spectral overlap density is

$$S_{DA} = \int \mathrm{g}_D(\mathrm{E}) \mathrm{g}_A(\mathrm{E}) \; dE$$

Since both line shape functions have units of inverse energy, $S_{DA}$ has $(\text{energy})^{-1}$. Within the weak-coupling limit, Fermi's golden rule gives the resonance energy transfer rate as,

$$k_{ET} = \frac{2\pi}{\hbar} V_{cpl}^2 S_{DA} \quad (10)$$

where ħ is the reduced Planck constant. Substitution of (9) into (10) gives;

$$k_{ET} = \frac{\rho_A \pi}{4z^4} \frac{1}{n^4} \left[\frac{\mu_A^2 \, \mu_{coh}^2}{8\pi\varepsilon_0^2 \hbar} S_{DA}\right] \quad (11)$$

which remains the plane-to-point distance dependance:

$$\boxed{k_{ET} \propto z^{-4}}$$

Experimentally we use;

$$J_{Spec} = \int \mathrm{F}_D(\lambda) \varepsilon_A(\lambda) \lambda^4 \; d\lambda \quad (12)$$

to calculate the spectral overlap integral ($J_{Spec}$), where, $\varepsilon_A(\lambda)$ is the absolute molar extinction coefficient so its magnitude contains the acceptor transition strength associated with $\mu_A^2$.

However, since the donor fluorescence spectrum $F_D(\lambda)$ is normalized $\int F_D(\lambda)\, d\lambda = 1$, its absolute transition strength is no longer contained in the spectrum. That information is introduced by the donor radiation rate;

$$k_{rad,D} = \frac{\Phi_D}{\tau_D}$$

Thus,

$$k_{ET} = \frac{\rho_A \pi}{4z^4} \frac{1}{n^4} \left[\frac{1}{8\pi\varepsilon_0^2 \hbar} \left(\frac{\Phi_D}{\tau_D}\right) J_{Spec}\right] \quad (13)$$

and finally,

$$\boxed{k_{ET} = 8.77 \times 10^{-11} \rho_A \left(\frac{\Phi_D}{\tau_D}\right) \frac{J_{Spec}}{n^4} \frac{\pi}{4z^4}} \quad (14)$$

Taking into account the combined ATTO 590 loading on both the top and bottom surfaces of the silica-encapsulated TDBC J-aggregate sheets of area A, the acceptor areal number density, $\rho_A$, for the construct with an experimentally determined ATTO 590:TDBC molecular ratio of 1:19 was calculated as:

$$\rho_A = \frac{N_{ATTO\ 590}}{A} = \frac{N_{ATTO\ 590}}{N_{TDBC}} \frac{N_{TDBC}}{A} = \frac{1}{19} \rho_{TDBC}$$

The molecular model of TDBC J-aggregates gives an approximate TDBC dimensions L = 2.0 nm, while the experimentally inferred lower bound for inter monomer plane spacing D = 0.7 nm.[4] Using this structural model as an estimate, the projected area occupied by one TDBC molecule in the J-agrgegate plane,

$$a_{TDBC} = LD = (2.0 \times 0.7)\ \text{nm}^2 = 1.4\ \text{nm}^2$$

$$\rho_{TDBC} = \frac{N_{TDBC}}{A} = \frac{1}{a_{TDBC}} = 0.71\, nm^{-2}$$

$$\rho_A = \frac{\rho_{TDBC}}{19} = 0.038\, nm^{-2}$$

With $\rho_A = 0.038$ nm$^{-2}$, J = 3.57 x $10^{15}$ M$^{-1}$ cm$^{-1}$ nm$^4$, z =3.9 nm, $\Phi_D = 0.94$, $\tau_D = 207$ ps, and n = 1.4 (accounted for mixed silica/aqueous surroundings as $n_{water} = 1.33$, $n_{silica} = 1.45$); using Eq (14) energy transfer rate is calculated: **$k_{ET} = 4.77 \times 10^{10} s^{-1}$**

The intrinsic donor ($k_D$) decay is,

$$k_D = \frac{1}{\tau_D} = \frac{1}{207\, ps} = 4.83\ \times 10^9\ s^{-1}$$

and the estimated energy transfer efficiency is,

$$E_{ET} = \frac{k_{ET}}{k_{ET}+k_D} = \mathbf{91\%} \quad (15)$$

The calculated energy-transfer rate is approximately ten times faster than the intrinsic donor decay of the TDBC J-aggregate exciton, indicating that energy transfer can effectively compete with the native excited-state relaxation of the J-aggregate.

## 8. References

1. Lakowicz, J. R. *Principles of Fluorescence Spectroscopy*, 3rd edn (Springer, 2006).

2. Chuang, C., Knoester, J. & Cao, J. Scaling relations and optimization of excitonic energy transfer rates between one-dimensional molecular aggregates. *J. Phys. Chem. B* **118**, 7827-7834 (2014).

3. Cao, J. Generalized resonance energy transfer theory: applications to vibrational energy flow in optical cavities. *J. Phys. Chem. Lett.* **13**, 10943-10951 (2022).

4. Thanippuli Arachchi, D. H. *et al.* Bright and fast emission from robust supramolecular J-aggregate nanostructures through silica-encapsulation. *ACS Nano* **18**, 20422-20434 (2024).